\documentstyle[preprint,aps,floats,epsfig]{revtex}

\def\appendix{\par
 \setcounter{section}{0}
 \setcounter{subsection}{0}
 \def\thesection{Appendix \Alph{section}}
 \def\theequation{\Alph{section}.\arabic{equation}}
 \setcounter{equation}{0}}
\begin{document}
\tightenlines
\renewcommand{\thefootnote}{\fnsymbol{footnote}}

\title{Nonperturbative dispersive sector in strong 
(quasi-)Abelian fields}

\author{G. Cveti\v c$^{a}$\footnote{
e-mail: cvetic@apctp.org; address after Sept.~15, 2000: Dept.~of Physics,
UTFSM, Valpara\'{\i}so, Chile}
and Ji-Young Yu$^{b}$\footnote{
e-mail: yu@dilbert.physik.uni-dortmund.de}
}

\address{ $^a$Asia Pacific Center for Theoretical Physics, 
Seoul 130-012, Korea}
\address{ $^b$Department of Physics, Dortmund University,
44221 Dortmund, Germany}

%

\maketitle

\renewcommand{\thefootnote}{\arabic{footnote}}

\begin{abstract}

In strong (quasi--)Abelian fields, even at the one--loop
level of the coupling constant, quantum fluctuations
of fermions induce an effective Lagrangian density
whose imaginary (absorptive) part is purely nonperturbative 
and known to be responsible for the fermion--antifermion 
pair creation. On the other hand, the induced real 
(dispersive) part has perturbative and nonperturbative
contributions. In the one--loop case, we argue how to separate 
the two contributions from each other for any strength of
the field. We show numerically that the nonperturbative
contributions are in general comparable with or larger
than the induced perturbative ones. We arrive at qualitatively
similar conclusions also for the induced energy density. 
Further, we investigate numerically the 
quasianalytic continuation of the perturbative results 
into the nonperturbative sector, by employing (modified) 
Borel--Pad\'e. It turns out that in the case at hand, 
we have to integrate over renormalon singularities, but
there is no renormalon ambiguity involved.\\
PACS number(s): 11.15.Bt, 10.10.Jj, 11.15.Tk, 11.80.Fv, 12.20.-m 

\end{abstract}

\section{Introduction}

It has been well known for some time 
that the effects of the
fermionic quantum fluctuations in space--time
uniform Abelian gauge fields can be effectively
integrated out, resulting in a one--loop
effective action 
\cite{Heisenberg:1936qt}--\cite{Itzykson:1980rh}. The results
have been formulated also for the covariantly
homogeneous, thus quasi--Abelian, fields of the $SU(2)$
gauge group \cite{Batalin:1977uv}, and for
specific nonhomogeneous magnetic field 
configurations \cite{Ragazzon:1995aw}. 
All the results can be extended to the case
of the quantum fluctuations of scalar particles.
The problems arising when genuinely non--Abelian
fields with translationally invariant gauge--invariants
are present [e.g., in $SU(3)_c$]
were discussed, e.g., in Refs.~\cite{Schwab:1982fy,Migdal:1985yu}.

The quantum fluctuations of the strong gauge field itself 
(photons, or gluons) modify 
additionally those Lagrangian densities 
induced by the fermionic quantum fluctuations.
Such two-loop effects have been successfully derived
by Ritus \cite{Ritus:1975cf} for homogeneous Abelian fields,
and further discussed by others 
\cite{Reuter:1996zm,Dunne:1999vd}.
In QED, such two--loop effects in the coupling constant
change the one--loop result by at most a few per cent
($ \pi \alpha$). We will omit them
in our investigation.

There are basically two classes of phenomena associated
with the presence of intense gauge fields.

Firstly, they can produce pairs of particles. 
For the Abelian case of strong homogeneous
electric field, Sauter \cite{Sauter} showed this
by investigating solutions of the Dirac equation
in the corresponding potential,\footnote{
A related problem was first considered 
even earlier by Klein \cite{Klein} 
who investigated solutions of the Dirac equation
with a high vertical barrier potential (Klein's paradox).}
and Schwinger \cite{Schwinger:1951ez} by using methods of action
integral, Green's functions and proper time.
Differential probabilities for pair creation
were investigated in Refs.~\cite{Nikishov}
and \cite{Casheretal}. In the latter
Reference, the quasi--Abelian model
was applied to investigation of the
quark pair production in chromoelectric flux tubes.
Experimental evidence related with the
pair production in a strong QED (laser) field
was reported in Ref.~\cite{Burke:1997ew}.

The pair production has its origin in the
imaginary (absorptive) part of the effective
Lagrangian induced by the fermionic quantum
fluctuations in the strong field. That part is
entirely nonperturbative in nature, because
the production effects are $\sim\!\exp(-{\rm const.\/}/g)$
and thus cannot be expanded in positive powers of the
field--to--fermion coupling parameter $g$. 

On the other hand, the other class of phenomena
is associated with the real (dispersive) part 
of the induced effective Lagrangian.
In QED, this class includes the following phenomena
that affect a low energy ($\omega \ll m_e$) photon wave 
entering the region of the strong background field:
photon splitting, change of the photon speed, and 
birefringence. Works on the theoretical aspects of these 
phenomena include Refs.~\cite{BB}--\cite{Dittrich:1998fy}. 
The experimental
aspects of birefringence in strong magnetic fields
are discussed in \cite{Bakalovetal}--\cite{Lee:1998ah}.
The dispersive part of the induced action leads
in principle to those corrections of the classical
Maxwell equations which originate from the (fermionic)
quantum fluctuations.

The aim of the present paper, while
dealing with the dispersive part of the
induced action, is somewhat different from these works.
We concentrate on the concept of separating
the nonperturbative from the perturbative
contributions in the induced dispersive action
when the product of the (quasi)electric field 
${\cal {E}}$ and the coupling constant $g$ is
large: $g {\cal {E}}/m^2 \stackrel{>}{\sim} 1$,
where $m$ is the fermion mass. Subsequently,
we numerically investigate the two contributions.
Afterwards, we use the discussed quantities
as a ``laboratory'' for testing and investigating the 
efficiency of methods of quasianalytic continuation.
The latter methods, involving the (modified) Borel-Pad\'e 
approximants, allow us to obtain approximately
the nonperturbative contributions from the 
approximate knowledge of the perturbative contributions 
and by employing the Cauchy principal value prescription
in the inverse Borel transformation (Laplace--Borel integral).
These considerations can give us insights into the
problems of extraction of nonperturbative
physics from the knowledge of perturbative
physics in gauge theories, in particular
in various versions of QCD (high--flavor, low--flavor).    

In Section II, we argue how to perform the
mentioned separation into the perturbative and
nonperturbative contributions. After identifying the 
two contributions,
we investigate numerically their values for various
values of the field parameter 
${\tilde a} \sim g {\cal {E}}/m^2$.
In Section III we then carry out an analogous analysis
for the induced energy density, 
the latter being in principle observable. 
In Section IV we then numerically investigate,
for the induced Lagrangian and energy densities,
(quasi)analytic continuation from the
perturbative into the nonperturbative sectors,
employing the method of Borel--Pad\'e for
the induced Lagrangian and a modified Borel--Pad\'e
for the induced energy density. 
We encounter integration over renormalon poles,
whose origin is nonperturbative, and we show
how to carry it out.
Section V summarizes our results and conclusions.

\section{Induced dispersive Lagrangian density}

We start by considering the Euler--Heisenberg Lagrangian density
\cite{Heisenberg:1936qt}--\cite{Itzykson:1980rh}
which is the part induced by the quantum
fluctuations of the fermions in an
arbitrarily strong (quasi--)Abelian
homogeneous field
\begin{eqnarray}
\delta {\cal L} & = & \frac{1}{8 \pi^2} 
\int_{- {\rm i} {\epsilon}}^{\infty - {\rm i} {\epsilon}} 
\frac{ds}{s} \exp[ - {\rm i} s ( m^2\!-\!{\rm i} {\varepsilon}^{\prime}) ]
\left[ g^2 a b \coth(a g s)  \cot(b g s) 
- \frac{g^2}{3}(a^2\!-\!b^2) - 1/s^2 \right] \ .
\label{EH1}
\end{eqnarray}
Here, $g$ is the field--to--fermion coupling parameter
(in electromagnetism it is the positron charge $e_0$), 
$m$ is the mass of the (lightest) fermion, 
and the parameters $a$ and $b$ are 
Lorentz--invariant expressions characterizing the (quasi)electric
and the (quasi)magnetic fields ${\vec {\cal E}}$
and ${\vec {\cal B}}$, respectively
\begin{eqnarray}
a & = & \left[ + {\vec {\cal E}}^2 - {\vec {\cal B}}^2
+ \sqrt{ \left( {\vec {\cal E}}^2 - {\vec {\cal B}}^2 \right)^2
+ 4 \left( {\vec {\cal E}} \cdot {\vec {\cal B}} \right)^2 }
\right]^{1/2}/\sqrt{2} \ ,
\label{a}
\\
b & = & \left[ - {\vec {\cal E}}^2 + {\vec {\cal B}}^2
+ \sqrt{ \left( {\vec {\cal E}}^2 - {\vec {\cal B}}^2 \right)^2
+ 4 \left( {\vec {\cal E}} \cdot {\vec {\cal B}} \right)^2 }
\right]^{1/2}/\sqrt{2} \ .
\label{b}
\end{eqnarray}
We note that 
$a b\!=\!|{\vec {\cal E}}\!\cdot\!{\vec {\cal B}}|$, and 
$a^2\!-\!b^2\!=\!{\vec {\cal E}}^2-{\vec {\cal B}}^2$.
Further, $a\!\to\!|{\vec {\cal E}}|$ 
when $|{\vec {\cal B}}|\!\to\!0$, 
and $b\!\to\!|{\vec {\cal B}}|$ 
when $|{\vec {\cal E}}|\!\to\!0$.
In the Lorentz frame where 
${\vec {\cal B}} \| {\vec {\cal E}}$, we simply have:
$a\!=\!|{\vec {\cal E}}|$ and $b\!=\!|{\vec {\cal B}}|$.
Expression (\ref{EH1}) can be obtained, for example, directly by
integrating out the fermionic degreees of freedom in the path
integral expression of the full effective action, and employing
the proper--time integral representation for the difference
of logarithms. 

As denoted, the integration in (\ref{EH1})
is performed along the positive real axis infinitesimally
below it, avoiding in this way the poles on the real axis 
appearing due to the (Lorentz--invariant version of the) 
magnetic field $b$ (function $\cot(b g s)$ there). 
If the path in (\ref{EH1}) were above the real axis, then we 
would obtain a nonzero imaginary part of the Lagrangian density 
even in the pure magnetic field case. This would imply particle 
creation in this case, which is physically unacceptable.
The path in (\ref{EH1}) reproduces for the case of the pure 
magnetic field the known real density, and for the case of
the pure electric field the known complex density
\cite{Schwinger:1951ez}. The path in (\ref{EH1}) 
is suggested also from the extension of the formal
approach of Ref.~\cite{Itzykson:1980rh}
to the general $a b\!\not=\!0$ case. 
Namely, when $a b\!\not=\!0$, 
we need to evaluate in this approach two traces:
one trace originating from $a\!\not=\!0$ and discussed in
\cite{Itzykson:1980rh} [their Eq.~(4-116)];
and the other trace
of the evolution operator of an harmonic oscillator,
originating from $b\!\not=\!0$ \cite{pc},
of the form $\sum \exp[- 2 {\rm i} s g b (n\!+\!1/2)]$ 
($n\!=\!0, 1, \ldots$; $s > 0$). The latter trace becomes
convergent after regularization 
$s \mapsto s\!-\!{\rm i}{\epsilon}$ (${\epsilon}\!\to\!+0$),
i.e., the path in (\ref{EH1}).

\begin{figure}[htb]
\setlength{\unitlength}{1.cm}
\begin{center}
\epsfig{file=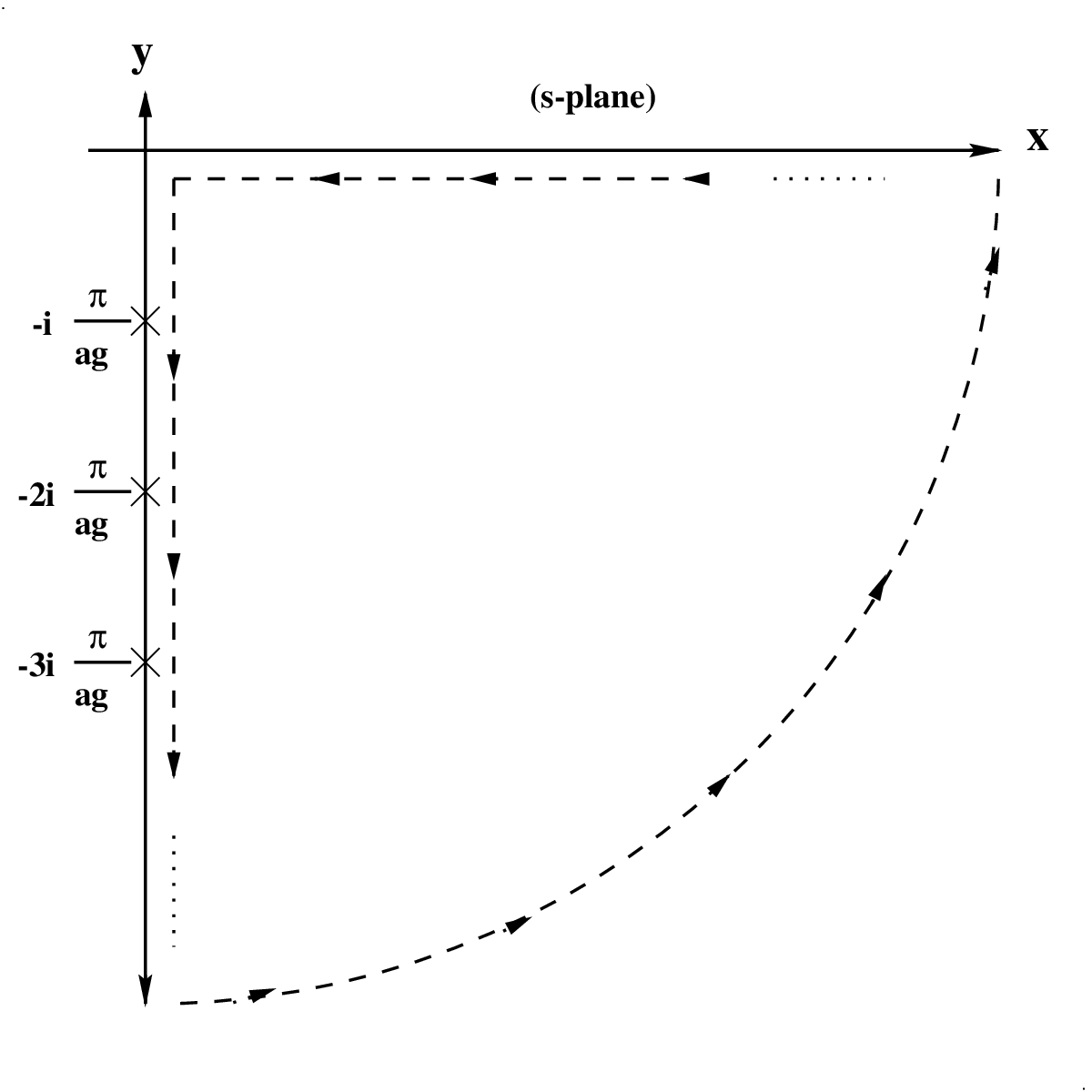, width=10.cm}
\end{center}
\vspace{-0.0cm}
\caption{\footnotesize
The contour integration, in the complex $s$--plane,
needed to rewrite (\ref{EH1}) in the form (\ref{EH2}).
The location of the poles is denoted explicitly.} 
\label{contour}
\end{figure}

Performing a contour integration in the fourth quadrant of the
complex proper--time $s$--plane (cf.~Fig.~\ref{contour}), 
expression (\ref{EH1}) can be rewritten as
\begin{eqnarray}
\delta {\cal L} & = & - \frac{1}{8 \pi^2} 
\int_{+0}^{\infty} 
\frac{dz}{(z\!+\!{\rm i}{\epsilon})} {\rm e}^{ - z m^2} 
\left[ g^2 a b \cot \left( a g ( z\!+\!{\rm i}{\epsilon} ) \right)  
\coth \left( b g ( z\!+\!{\rm i}{\epsilon} ) \right) 
+ \frac{g^2}{3} (a^2\!-\!b^2)
- \frac{1}{  ( z\!+\!{\rm i}{\epsilon} )^2 } \right] ,   
\nonumber\\
&&
\label{EH2}
\end{eqnarray}
where $s\!=\!- {\rm i} z\!+\!{\epsilon}$ now runs
along the negative imaginary axis (and $\epsilon\!\to\!+0$). 
As seen from Fig.~\ref{contour}, the result (\ref{EH2})
is actually independent of the precise path of the proper time
variable $s$ in (\ref{EH1}), as long as $s$ runs from near the
origin towards $s\!=\!+\infty$ and passes each positive pole
on the right, i.e. precisely the class of paths satisfying the
physical condition that the pure magnetic fields cannot
produce particles.
In (\ref{EH1}) and (\ref{EH2}), 
the familiar \cite{Heisenberg:1936qt} 
counterterm $\propto\!(a^2\!-\!b^2)$ 
[$=\!({\cal E}^2\!-\!{\cal B}^2)$]
is included which makes the integral finite. 
This divergent term leads to the renormalization
of the field in the leading Lagrangian density 
${\cal L}^{(0)} = ({\cal E}^2\!-\!{\cal B}^2)/2$.
We now divide the integration region into intervals for the
integration variable $a g z$: $i_0\!=\![0, \pi/2]$,
$i_1\!=\![\pi/2, 3 \pi/2]$, $\ldots$, 
$i_n\!=\![ (n\!-\!1/2)\pi,(n\!+\!1/2)\pi]$, $\ldots$
Each interval, except $i_0$, contains in its middle
one pole of the integrand. The corresponding series
for the real (dispersive) part of the Lagrangian density
is
\begin{eqnarray}
{\rm Re} \delta {\tilde {\cal L}} &=& 
{\rm Re} \delta {\tilde {\cal L}}_0 + 
\sum_{n=1}^{\infty} {\rm Re} \delta {\tilde {\cal L}}_n \ ,
\label{sum}
\\
{\rm Re} \delta {\tilde {\cal L}}_0 &=& 
- \int_0^{\pi/2}
\frac{d w}{w} \exp \left(- \frac{w}{\tilde a} \right) 
\left[ p \cot (w) \coth(p w)
+ \frac{1}{3} (1\!-\!p^2) - \frac{1}{w^2} \right] \ ,
\label{L0}
\\
{\rm Re} \delta {\tilde {\cal L}}_n &=& 
- \exp \left( - \frac{n \pi}{{\tilde a}} \right) 
{\Bigg \{} \int_{-\pi/2}^{\pi/2} dw 
\exp \left( - \frac{w}{\tilde a} \right) 
{\Bigg [} \frac{p \cot (w) \coth(p (w\!+\!n \pi)) }{(w\!+\!n \pi)} 
\nonumber\\
&&- \frac{p}{w} \frac{\coth( p n \pi)}{ n \pi} 
+ \frac{(1\!-\!p^2)}{3 (w\!+\!n \pi)} -
\frac{1}{(w\!+\!n \pi)^3} {\Bigg ]}
\nonumber\\
&&  + {\rm Re} \int_{-\pi/2}^{\pi/2} dw 
\exp \left( - \frac{w}{\tilde a} \right) 
\frac{1}{ (w\!+\!{\rm i} {\epsilon}^{\prime}) } 
\frac{p \coth( p n \pi)}{ n \pi} {\Bigg\}} \ , \quad (n\!\geq\!1) \ .
\label{Ln}
\end{eqnarray}
Here, ${\epsilon}^{\prime}\!\equiv\!{\epsilon} g a \to +0$, 
and we used the notation
\begin{equation}
{\tilde a}\equiv\frac{g a}{m^2}, \ {\tilde b}\equiv\frac{g b}{m^2}, 
\qquad p \equiv \frac{b}{a} \equiv 
\frac{{\tilde b}}{{\tilde a}} \ , \qquad  
\delta {\tilde {\cal L}} \equiv 
\delta {\cal L}/ \left( \frac{m^4 {\tilde a}^2}{8 \pi^2} \right) \ ,
\label{not}
\end{equation}
and we introduced the dimensionless integration variable 
$w\!\equiv\!agz$ when $agz$ is in the interval $i_0$, and 
$w\!\equiv\!agz\!-\!n \pi$ when $agz$ 
is in the interval $i_n$ ($n\!\geq\!1$).
In (\ref{Ln}), we separated the integrand into a part that is
entirely nonsingular in the integration region, and a part that
is singular but gives a finite value of integration 
since the Cauchy principal (${\cal P}$) value has to be taken. 
{}From a formal point of view, we note that
$\delta {\tilde {\cal L}}_0$ doesn't ``feel'' the poles
of the integrand as depicted in Fig.~\ref{contour},
while $\delta {\tilde {\cal L}}_n$ ($n\!\geq\!1$)
``feels'' the pole $s\!=\!-{\rm i} n \pi/(ag)$
via the principal value part in (\ref{Ln}) that is
proportional to
\begin{eqnarray}
 {\rm Re} \int_{-\pi/2}^{\pi/2}\!dw 
\frac{ \exp ( - w/{\tilde a} ) }
{ (w\!+\!{\rm i} {\epsilon}^{\prime}) } &\equiv&
{\cal P} \int_{-\pi/2}^{\pi/2}\!\frac{dw}{w} 
\exp ( - w/{\tilde a} ) =
\nonumber\\
- {\rm E}_1 \left(\frac{\pi}{2 {\tilde a}}\right)\!-\!{\rm Ei}
\left(\frac{\pi}{2 {\tilde a}}\right)
&=& \left \{
\begin{array}{l l}
- 2 \left[ x\!+\!x^3/(3!\,3)\!+\!x^5/(5!\,5)\!+\!
\cdots \right] |_{x=\pi/(2 {\tilde a})} & 
\text{if ${\tilde a} \gg 1$} \ , \\
 - (e^x/x) \left[ 1\!+\!(1/x)\!+\!
\cdots \right] |_{x=\pi/(2 {\tilde a})} &
\text{if ${\tilde a} \ll 1$} \ .
\end{array}
\right \}
\label{princ}
\end{eqnarray} 
We note that the dispersive part of the induced Lagrangian density 
as normalized here (\ref{sum})--(\ref{not}) 
depends only on two dimensionless parameters --
on parameter $p\!\equiv\!b/a$ which characterizes 
in a Lorentz--invariant manner
the ratio of the strengths of the (quasi)electric and
(quasi)magnetic fields [cf.~(\ref{a})--(\ref{b})],
and on parameter ${\tilde a}\!\equiv\!(g a)/m^2$ 
which characterizes the combined strengths of 
the (quasi)electric field parameter $a$
and the field--to--fermion coupling $g$. In the perturbative
weak--field limit, ${\tilde a}$ is small. 
In this case, when reintroducing
in (\ref{L0}) $z\!\equiv\!w/(ag)$
\begin{eqnarray}
{\rm Re} \delta {\cal L}_0 &=& 
 - \frac{1}{8 \pi^2} 
\int_{0}^{\pi/(2 a g)} 
\frac{dz}{z} \exp( - z m^2) 
\left[ g^2 a b \cot ( a g z ) \coth ( b g  z) +
\frac{g^2}{3}(a^2\!-\!b^2) - \frac{1}{z^2}\right],
\nonumber\\
&&
\label{EH3}
\end{eqnarray}
we can see that the real part of
expression (\ref{EH2}) is approximately reproduced,
since formally $\pi/(2 a g)\!\to\!\infty$.
In this case, the conventional perturbative
expansion of the dispersive Lagrangian density in
powers of $g^2$ (i.e., inverse powers of $x$)
can be performed (cf.~\cite{Heisenberg:1936qt}, 
\cite{Schwinger:1951ez})
\begin{equation}
\delta {\tilde {\cal L}}^{\rm pert.} =
\left(
c_1 1!\,{\tilde a}^2 + c_3 3!\,{\tilde a}^4 + 
c_5 5!\,{\tilde a}^6 + \cdots
\right) \ ,
\label{Lpert}
\end{equation}
where the expansion coefficients are
\begin{eqnarray}
c_1 & = &  \frac{1}{45} \left[ 
(1\!-\!p^2)^2 + 7 p^2 \right] \ , \quad
c_3 =  \frac{1}{945} \left[ 
2 (1\!-\!p^2)^3 + 13 p^2 (1\!-\!p^2) \right] \ ,  
\nonumber\\
c_5 & = & \frac{1}{14175} \left[
3 (1\!-\!p^2)^4 + 22 p^2 (1\!-\!p^2)^2 + 19 p^4 
\right] \ , \ {\rm etc.}
\label{cjs}
\end{eqnarray}
Expression (\ref{Lpert}) can be derived alternatively by
purely perturbative methods --
the terms $\sim$${\tilde a}^{2 n}$ can be obtained
by calculating the one--fermion--loop Feynman diagram 
with $2 n$ photon external legs of zero momenta.
Expression (\ref{Lpert}) is a divergent asymptotic series
and it gives the usual perturbative corrections
to the Maxwell equations. 
On the other hand, the formal small--${\tilde a}$ expansion of
${\rm Re} \delta {\tilde {\cal L}}_0$ of (\ref{L0}) 
[or equivalently: (\ref{EH3})] reproduces the
terms (\ref{Lpert}) and yields in addition the 
terms $\sim$${\tilde a} \exp(-{\rm const.\/}/{\tilde a})$.
The latter terms may in principle be dangerous for the interpretation
of ${\rm Re} \delta {\tilde {\cal L}}_0$ of (\ref{L0}) as
the perturbative part of the induced density, since they are
nonanalytic and could thus signal physical nonperturbative
effects. However, in the Appendix we demonstrate that
these terms are only an artifact of the abruptness of the
infrared (IR) proper--time cutoff 
$z\!\leq\!1/{\Lambda}^2_{\rm IR}$
(${\Lambda}^2_{\rm IR}\!=\!(2/\pi) 
m^2 {\tilde a}\!\sim\!m^2 {\tilde a}$).\footnote{
The energy cutoff
${\Lambda}_{\rm IR}\!\sim\!m \sqrt{{\tilde a}}$ 
is low in the case when the perturbative effects dominate
(i.e., at ${\tilde a}\!<\!1$), but is higher when the
nonperturbative effects are significant
(at ${\tilde a}\!>\!1$). 
The nonperturbative effects here
reside in the infrared (IR) sector
of (fermionic) momenta $q\!<\!{\Lambda}_{\rm IR}$, 
and the effective contributing size of this sector gets
larger when ${\tilde a}$ grows.}
These terms are therefore not of a physical nonperturbative
origin. In the Appendix we further show that
${\rm Re} \delta {\tilde {\cal L}}_0$ of (\ref{L0})
should be reinterpreted as the limit with an
infinitesimally softened IR cutoff, the latter
limit being numerically the same but having no 
nonanalytic terms in the small--${\tilde a}$
expansion. That expansion is then identical to (\ref{Lpert}).

On the other hand, the densities ${\rm Re} \delta {\tilde {\cal L}}_n$
($n\!\geq\!1$) of (\ref{Ln}) represent the nonperturbative part
of the induced dispersive density (\ref{sum}), for two reasons:
\begin{itemize}
\item
The integration over the proper--time
$z$ runs here in the vicinity (in fact, across) the
$n$'th pole of the integrand of (\ref{EH2}). The poles of the 
integrand are in the nonperturbative regions. We recall
that these poles are also the source of 
the nonzero imaginary (absorptive)
part of the density leading to the fermion--antifermion
pair creation, a clearly nonperturbative phenomenon.
\item
The densities
${\rm Re} \delta {\tilde {\cal L}}_n$ ($n\!\geq\!1$),
independently of the pole structure of their integrands,
become appreciable only in the strong--field (large--${\tilde a}$)
regime while in the weak--field (small--${\tilde a}$) regime 
they decrease faster than any power of ${\tilde a}$, i.e.,
they do not contribute to the perturbative series (\ref{Lpert}).
Each of the two integrals
in the curly brackets of (\ref{Ln}) behaves as
$\sim\!{\tilde a} \exp[\pi/(2 {\tilde a})]$
when ${\tilde a}\!\to\!+0$, and thus the
entire ${\rm Re} \delta {\tilde {\cal L}}_n$ 
behaves as 
$\sim\!{\tilde a}\exp[- (n\!-\!1/2) \pi/{\tilde a}]$
($n\!\geq\!1$) in this limit.\footnote{
If we did not take the principal Cauchy value
in (\ref{Ln}), but some other prescription
(which in the case at hand would be wrong),
${\rm Re} \delta {\tilde {\cal L}}_n$ would
behave as $\sim\!\exp[- n \pi/{\tilde a}]$.}
\end{itemize}

Therefore, the nonperturbative effects
contained in ${\rm Re} \delta {\tilde {\cal L}}_n$ ($n\!\geq\!1$)
are of two types, one type being characterized by
the poles of the integrand, and the other type by
what we may call strong--field effects. 
{}From the above discussion,
it further follows that we have some freedom in
choosing the IR proper--time cutoff: 
$z\!\leq\!1/{\Lambda}^2_{\rm IR}$ is such that
${\Lambda}^2_{\rm IR}\!\sim\!m^2 {\tilde a}$ and that
all the possible poles must lie at $z$'s above the
cutoff $1/{\Lambda}^2_{\rm IR}$. We took
${\Lambda}^2_{\rm IR}\!=\!\kappa m^2 {\tilde a}$
with $\kappa\!=\!2/\pi$, but any $\kappa$
satisfying $1/\pi\!<\!\kappa\!\sim\!1$
would be acceptable as well.

{}From a somewhat different perspective, we can
imagine transforming a truncated perturbation expansion for 
${\rm Re} \delta {\tilde {\cal L}}^{\rm pert.}/{\tilde a}$
of (\ref{Lpert}) (with several terms)
via the Borel--Pad\'e approximation. The resulting
integrand approximately reproduces the integrand of
(\ref{EH2}), including the poles structure. Thus
the integration over the $n$'th pole, contained in
${\rm Re} \delta {\tilde {\cal L}}_n$ of (\ref{Ln}),
can be interpreted as the $n$'th renormalon
in the density, i.e., a nonperturbative quantity.
We will return to this point later in this paper.

Thus, the densities (\ref{L0}) and (\ref{Ln})
result in the fermion--induced perturbative and nonperturbative
contributions, respectively, to the Maxwell
equations. The fields were
taken, strictly speaking, to be homogeneous in space and time.
In practical terms, this means that they are
not allowed to change significantly on the distance
and time scales of the Compton wavelength of the
fermion $1/m$. For electro--magnetic fields,
$m$ is the electron mass, and $1/m$ is about 
$4 \cdot 10^{-13}$ m, and $1.3 \cdot 10^{-21}$ s.

\begin{figure}[b]
\begin{minipage}[b]{.49\linewidth}
 \centering\epsfig{file=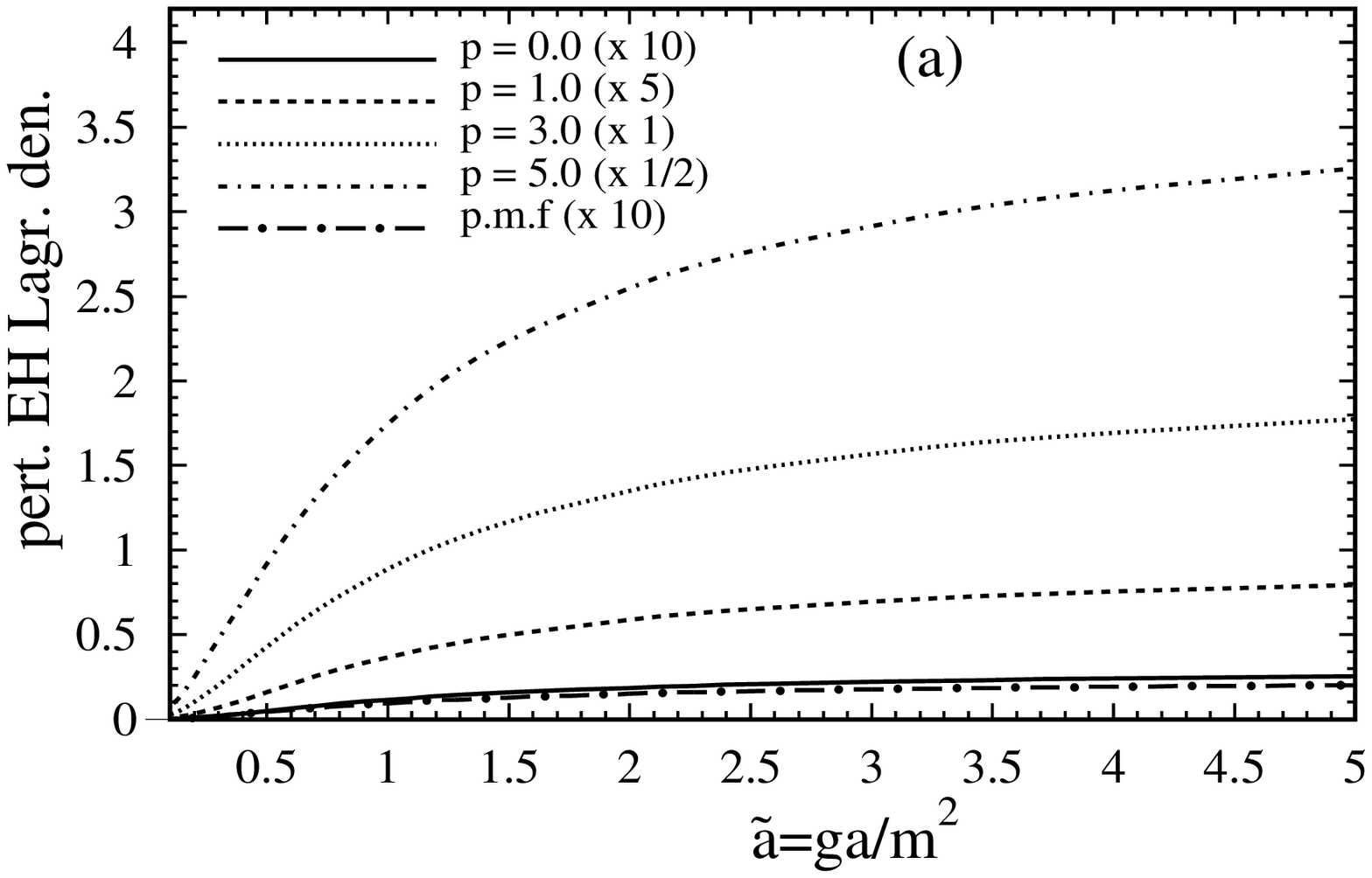,width=\linewidth}
\end{minipage}
\begin{minipage}[b]{.49\linewidth}
 \centering\epsfig{file=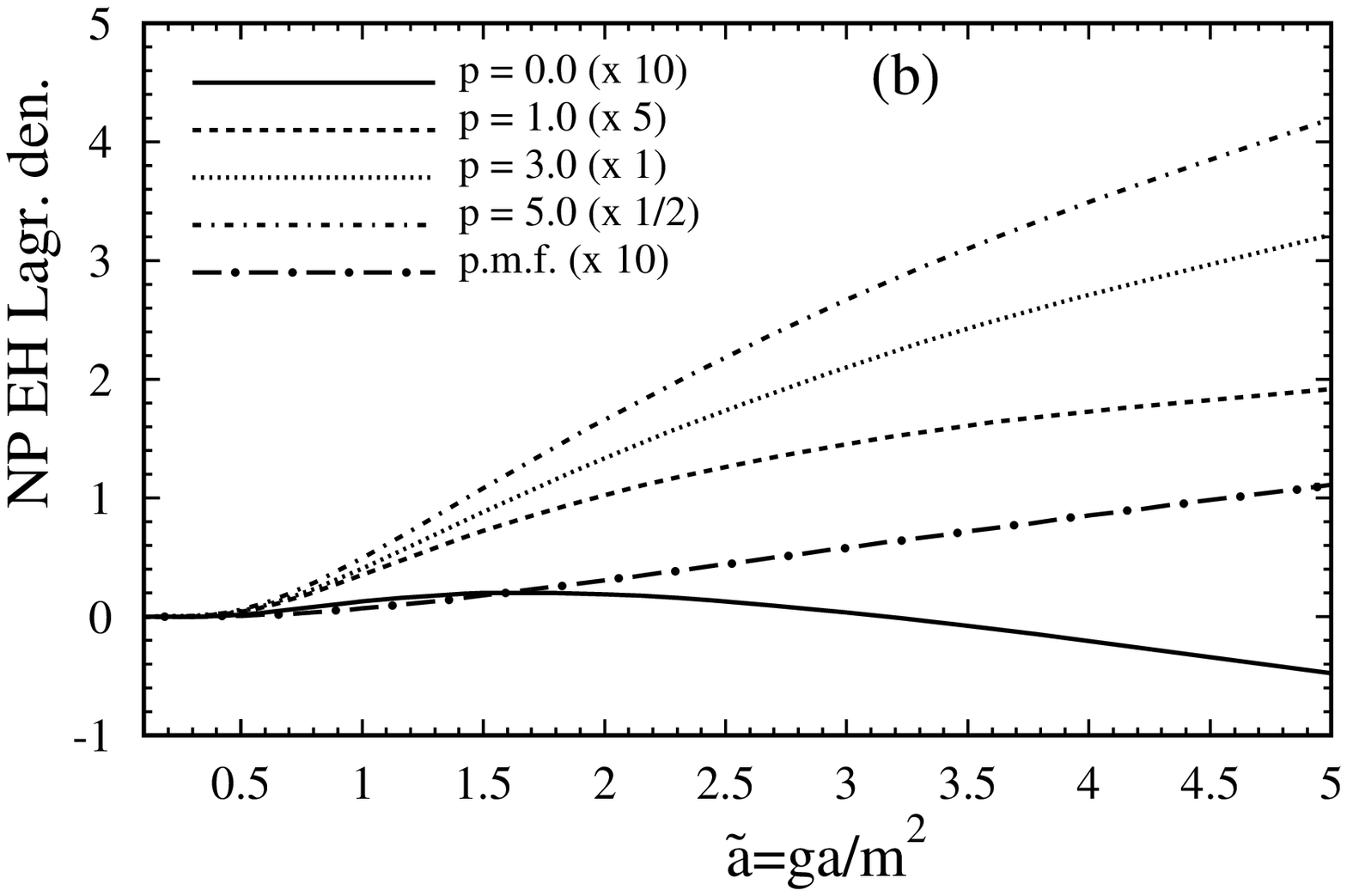,width=\linewidth}
\end{minipage}
\vspace{-1.0cm}
\caption{\footnotesize
(a) Perturbative and (b) nonperturbative
induced dispersive (Euler--Heisenberg) Lagrangian densities 
[cf.~(\ref{L0}) and (\ref{Ln})]
as functions of the (quasi)electric field
parameter ${\tilde a}$ (\ref{not}),
at various fixed values of the
magnetic--to--electric field ratio 
$p\!=\!{\tilde b}/{\tilde a}$ (\ref{not}).
The actual values of the curves for 
$p\!\approx\!0$, $p\!=1.0$ and $p\!=\!5.0$
have been multiplied here by factors
$10$, $5$ and $1/2$, respectively, for better visibility.
Included is also the case of the pure (quasi)magnetic field
(p.m.f.), for which the $x$-axis represents ${\tilde b}\!=\!g b/m^2$.}
\label{Lvsa}
\end{figure}

\begin{figure}[t]
\centering\epsfig{file=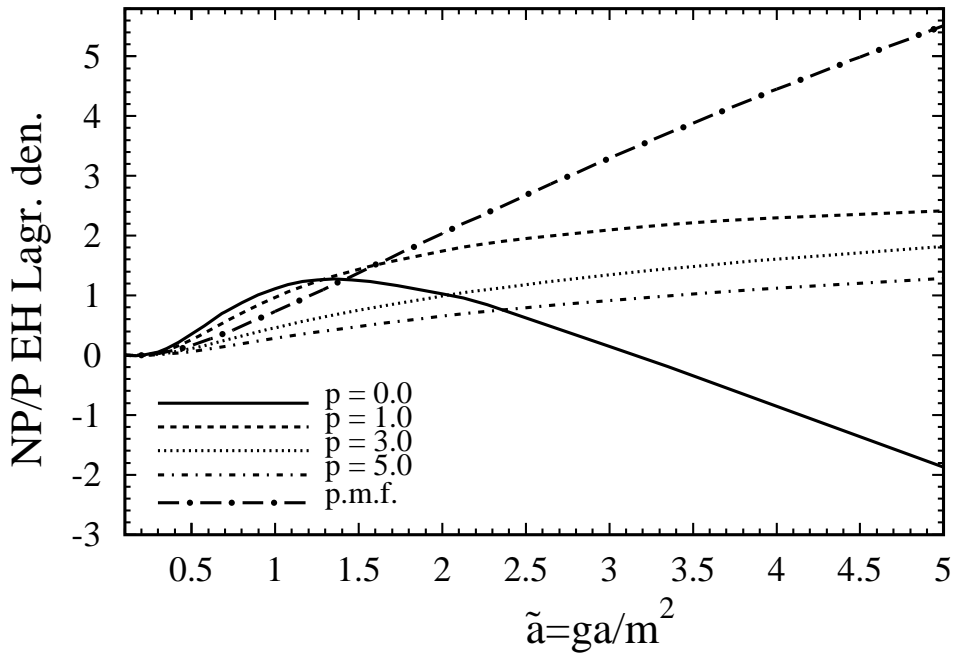,width=14.cm}
\vspace{0.2cm}
\caption{\footnotesize
Ratios of the nonperturbative and the
perturbative induced dispersive Lagrangian densities
for the cases depicted in Figs.~\ref{Lvsa}.
For the p.m.f. case,
the $x$-axis represents ${\tilde b}\!=\!g b/m^2$.}
\label{Lratio}
\end{figure}

An indication of the relative size of the 
perturbative and nonperturbative 
fermion--induced corrections to the Maxwell equations
can be obtained by comparing the corresponding
contributions to the induced Lagrangian density.
This is done in Figs.~\ref{Lvsa}-\ref{Lratio}.
Figures \ref{Lvsa} (a), (b) show the
dimensionless perturbative (\ref{L0})
and nonperturbative (\ref{Ln}) induced
Lagrangian densities, respectively,
as functions of the (quasi)electric
field parameter ${\tilde a}$ (\ref{not}), at four
different fixed values $p\!\equiv\!{\tilde b}/{\tilde a}$ 
of the magnetic--to--electric field ratio.
The case of the pure (quasi)magnetic field (p.m.f.) 
is also included in the Figures, as function of ${\tilde b}$.
For the p.m.f. case, we normalized the Lagrangian
density in analogy with (\ref{not}), i.e., 
$\delta {\tilde {\cal L}}$ is obtained in that case
by dividing $\delta {\cal L}$ by 
$m^4 {\tilde b}^2/(8 \pi^2)$. The separation
between the perturbative and the nonperturbative part
was performed in the p.m.f. case analogously,
i.e., the proper--time $z\!<\!\pi/(2 b g)$ contributions
were defined to be perturbative, and those
from $z\!>\!\pi/(2 b g)$ nonperturbative. We point
out, however, that in the latter case the nonperturbative
contributions do not involve the
renormalon--type (``pole--type'') effects, 
but only strong--field effects (cf.~previous discussion).
In Fig.~\ref{Lratio}, the corresponding
ratios of the nonperturbative and perturbative
induced densities are presented.\footnote{
The total induced dispersive
Lagrangian densities, and values of the truncated
perturbation series (\ref{Lpert})
(including $\sim\!{\tilde a}^8$), are included
in Figs.~\ref{LPade} in Section IV.}
When moving beyond the perturbative
region (i.e., when ${\tilde a}\!\not\ll\!1$),
we see from these Figures that the nonperturbative
parts in general become relatively significant
and often even dominant.

Once we come into the nonperturbative regime
(${\tilde a}\!\stackrel{>}{\sim}\!1$), 
however, we must keep in
mind that the pair creation, originating from the large
absorptive part, will become so strong as to render
the solutions of the corrected
Maxwell equations unstable. 
We will quantify somewhat this fact in the next Section
in the case of the induced energy density in QED.

In Figs. \ref{Lvsa} (a), (b), the densities were normalized
according to (\ref{not}), so that the tree--level
reference values for the densities are
\begin{equation}
{\tilde {\cal L}}^{(0)} \equiv 
{\cal L}^{(0)}/\left( \frac{m^4 {\tilde a}^2}{8 \pi^2} \right)
= \frac{4 \pi^2}{g^2} (1 - p^2) \ .
\label{treeL}
\end{equation}
Therefore, increasing only the coupling parameter $g$,
while leaving the (quasi)electric field $a$ unchanged,
results in correspondingly larger relative corrections
originating from the induced parts, both nonperturbative
and perturbative. In the special case of QED, 
on the other hand, $g\!=\!e_0$ is
small [$\alpha\!=\!e_0^2/(4 \pi)\!\approx\!1/137$], 
and the overall induced Lagrangian density accounts
usually for less than $0.5$ per mille of the total Lagrangian
density when ${\tilde a}\!\leq\!1$
(see the next Section on related points).

\section{Induced energy density}

\begin{figure}[b]
\begin{minipage}[b]{.49\linewidth}
 \centering\epsfig{file=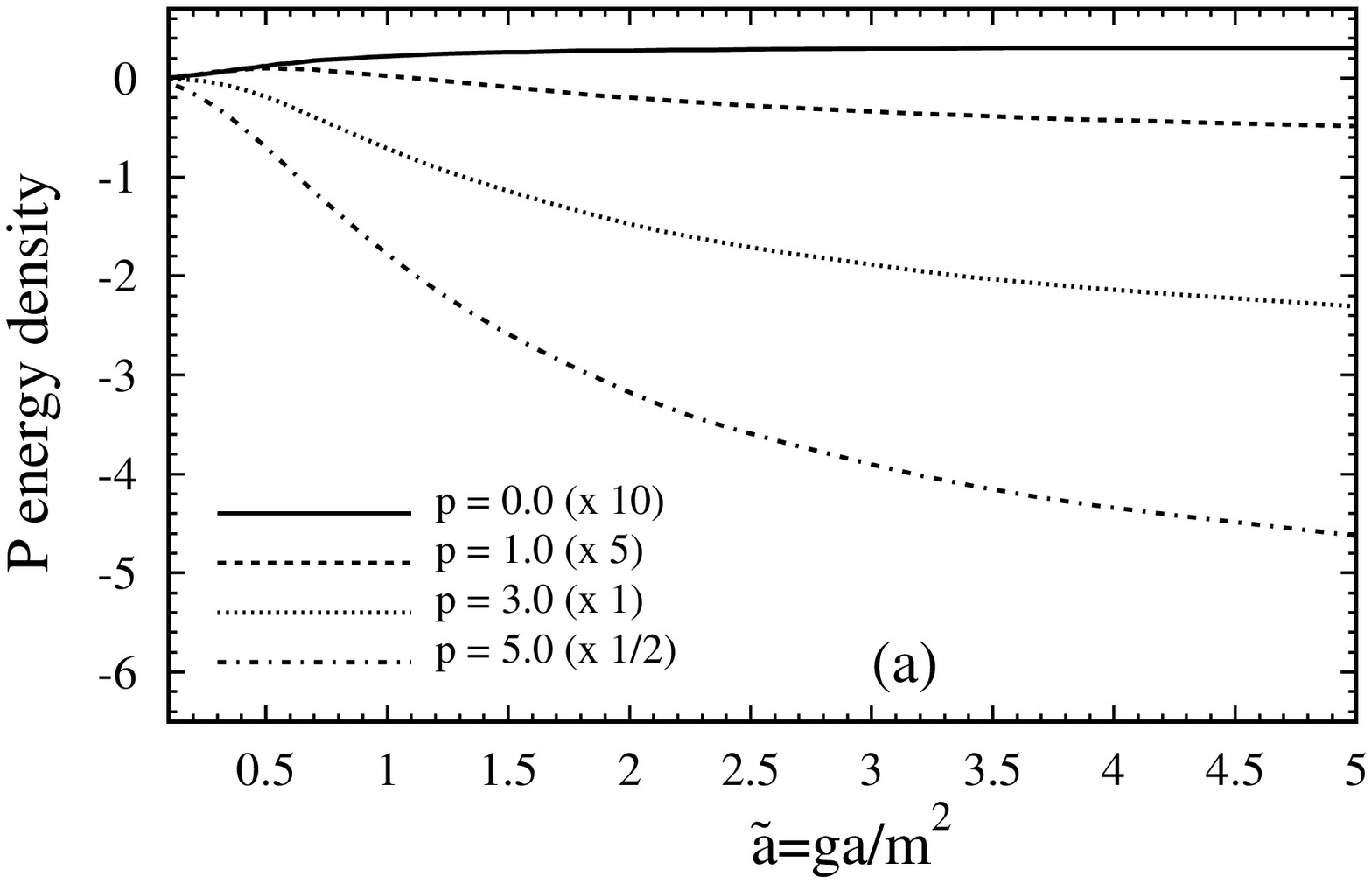,width=\linewidth}
\end{minipage}
\begin{minipage}[b]{.49\linewidth}
 \centering\epsfig{file=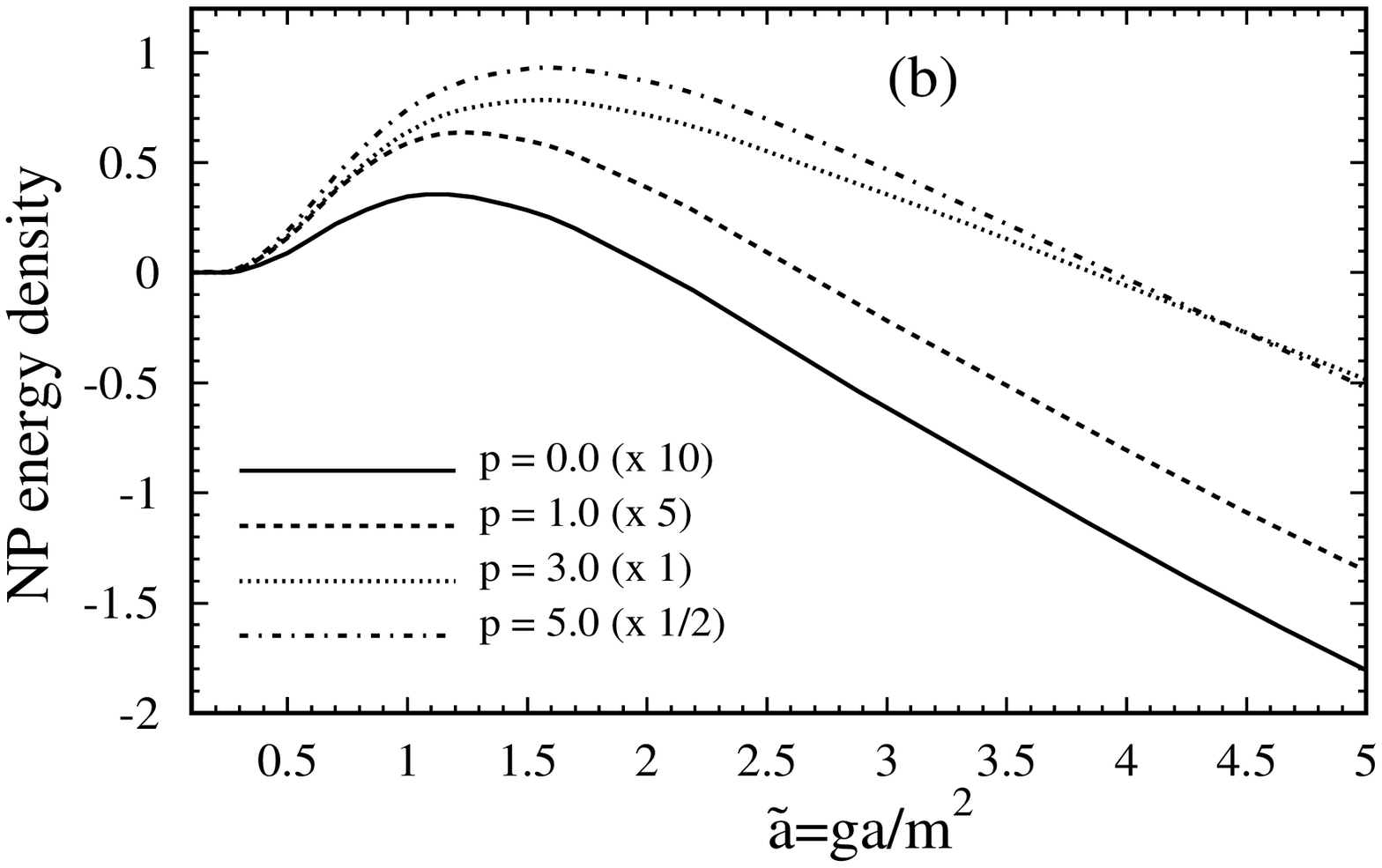,width=\linewidth}
\end{minipage}
\vspace{-1.0cm}
\caption{\footnotesize
(a) Perturbative and (b) nonperturbative
induced energy densities [cf.~(\ref{U0}) and (\ref{Un})]
as functions of ${\tilde a}$
at various fixed values of 
$p\!=\!{\tilde b}/{\tilde a}$ (\ref{not}).
The actual values of the curves have been
multiplied, for better visibility, by the denoted factors, 
just as in Figs.~\ref{Lvsa}.}
\label{Uvsa}
\end{figure}

\begin{figure}[t]
\setlength{\unitlength}{1.cm}
\begin{center}
\epsfig{file=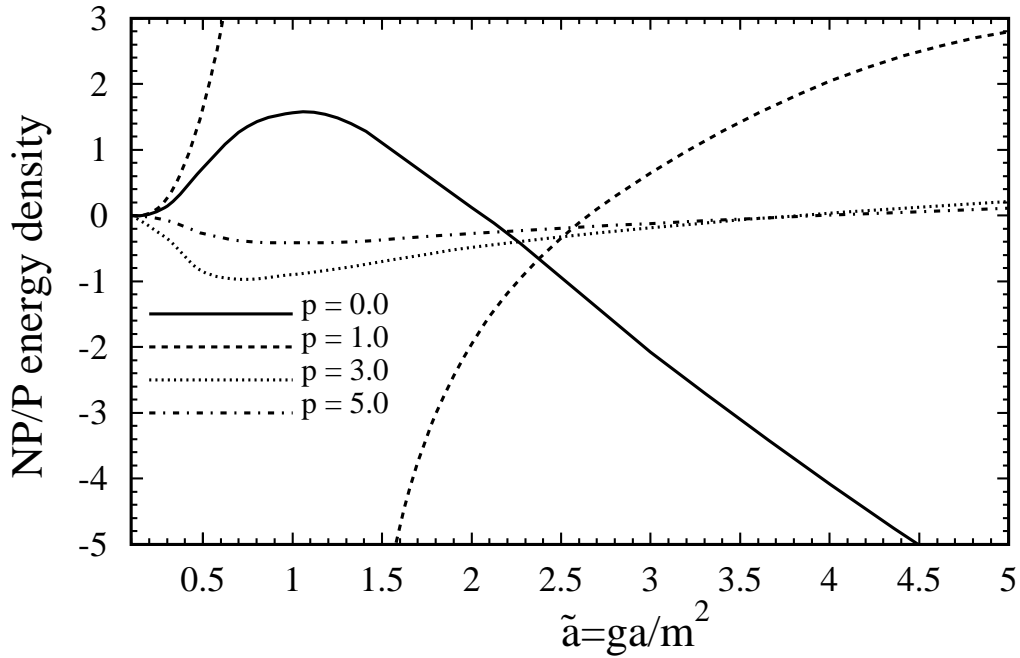,width=14.cm}
\end{center}
\vspace{-0.3cm}
\caption{\footnotesize
Ratios of the nonperturbative and the
perturbative induced energy densities
for the cases depicted in Figs.~\ref{Uvsa}.
The ratio for $p\!=\!1$ varies strongly for
${\tilde a}\!=\!0.5$--$1.5$ because the perturbative
induced density has a zero at ${\tilde a}\!\approx\!1.1$.}
\label{Uratio}
\end{figure}

In this Section, we discuss the induced energy densities.
Energy density is in principle a measurable quantity.
It is not Lorentz--invariant.
If the (quasi)electric and (quasi)magnetic fields
are mutually parallel, the various induced energy densities
can be obtained directly from the corresponding
induced Lagrangian densities
\begin{eqnarray}
\delta {\cal U}_k &=& 
a \frac{{\partial} {\rm Re} \delta {\cal L}_k}{{\partial} a}
{\Bigg |}_b 
- {\rm Re} \delta {\cal L}_k 
\quad 
\nonumber\\
\Rightarrow \quad
\delta {\tilde {\cal U}}_k &=& 
{\tilde a} \frac{{\partial} {\rm Re} \delta {\tilde {\cal L}}_k}
{{\partial} {\tilde a}} {\Bigg |}_{{\tilde b}} 
+ {\rm Re} \delta {\tilde {\cal L}}_k 
\quad  (k = 0, 1, 2, \ldots) \ ,
\label{delUk}
\end{eqnarray}
where we denoted, in analogy with (\ref{not})
\begin{equation}
\delta {\tilde {\cal U}}_{(k)} \equiv 
\delta {\cal U}_{(k)}/ 
\left( \frac{m^4 {\tilde a}^2}{8 \pi^2} \right) 
\ .
\label{notU}
\end{equation}
With the restriction to parallel fields 
${\vec {\cal E}} \parallel {\vec {\cal B}}$ 
(i.e, $|{\vec {\cal E}}|\!=\!a$
and  $|{\vec {\cal B}}|\!=\!b$) we do not lose the
generality since, for any configuration of ${\vec {\cal E}}$
and ${\vec {\cal B}}$, there always exists a Lorentz boost,
perpendicular to the plane of the fields, so that in the
boosted frame the two fields are parallel.
The corresponding perturbative and nonperturbative parts
of the energy densities in such frames are
\begin{eqnarray}
{\rm Re} \delta {\tilde {\cal U}}_0 &=& 
- \int_0^{\pi/2}
\frac{d w}{w} \exp \left(- \frac{w}{\tilde a} \right) 
{\Bigg\{} \left( \frac{w}{\tilde a} - 1 \right)
\left[ p \cot (w) \coth(p w)
+ \frac{1}{3} (1\!-\!p^2) - \frac{1}{w^2} \right] 
\nonumber\\
&&+ \left[  p \cot (w) \coth(p w) + p^2 \frac{ w \cot(w) }{ \sinh^2(p w)}
+ \frac{2}{3} - \frac{2}{w^2} \right] {\Bigg \}} \ ,
\label{U0}
\\
{\rm Re} \delta {\tilde {\cal U}}_n &=& 
- \exp\left(- \frac{n \pi}{{\tilde a}} \right) {\Bigg \{}
\int_{-\pi/2}^{\pi/2} dw \exp \left(- \frac{w}{\tilde a} \right)
\left[ \frac{(w\!+\!n \pi)}{\tilde a} - 1 \right]
{\Bigg [} \frac{p \cot (w) \coth (p (w\!+\!n \pi)) }{(w\!+\!n \pi)} 
\nonumber\\
&&- \frac{p}{w} \frac{\coth( p n \pi)}{ n \pi} 
+ \frac{(1\!-\!p^2)}{3 (w\!+\!n \pi)} -
\frac{1}{(w\!+\!n \pi)^3} {\Bigg ]}
\nonumber\\
&&+ \int_{-\pi/2}^{\pi/2} dw \exp\left(- \frac{w}{\tilde a} \right)
{\Bigg [} p \cot(w) \left( 
\frac{ \coth (p (w\!+\!n \pi))}{(w\!+\!n \pi)} +
\frac{p}{\sinh^2 (p (w\!+\!n \pi))} \right)
\nonumber\\
&&- \frac{p}{w} \left( \frac{ \coth(p n \pi)}{n \pi}
+ \frac{p}{\sinh^2 (p n \pi)} \right) + 
\frac{2}{3} \frac{1}{(w\!+\!n\pi)} - \frac{2}{(w\!+\!n\pi)^3}
{\Bigg ]}
\nonumber\\
&&+ \left[ \frac{p}{\tilde a} \coth( p n \pi) +
 \frac{p^2}{\sinh^2 (p n \pi)} \right] 
\left[ - {\rm E}_1 \left(\frac{\pi}{2 \tilde a} \right) -
 {\rm Ei}\left(\frac{\pi}{2 \tilde a} \right) \right]
\nonumber\\
&&+ \frac{2 p}{n \pi} \coth(p n \pi) 
\sinh \left( \frac{\pi}{2 \tilde a} \right)
{\Bigg \}} \ , \quad (n \geq 1) \ .
\label{Un}
\end{eqnarray}
The tree--level density in the normalization convention used 
[cf.~(\ref{not})] is
\begin{equation}
{\tilde {\cal U}}^{(0)} \equiv
{\cal U}^{(0)}/ \left( \frac{m^4 {\tilde a}^2}{8 \pi^2} \right)
= \frac{4 \pi^2}{g^2} (1 + p^2) \ .
\label{Utree}
\end{equation}
The perturbative power expansion
of the induced energy density $\delta {\tilde {\cal U}}$ is
\begin{equation}
\delta {\tilde {\cal U}}^{\rm pert.} =
\left(
d_1 1!\,{\tilde a}^2 + d_3 3!\,{\tilde a}^4 + 
d_5 5!\,{\tilde a}^6 + \cdots
\right) \ ,
\label{Upert}
\end{equation}
where the expansion coefficients are
\begin{eqnarray}
d_1 &=&  \frac{1}{45} \left[ 
3 + 5 p^2 - p^4 \right] \ , \quad
d_3 =  \frac{1}{945} \left[ 
10 + 21 p^2 - 7 p^4 + 2 p^6 \right] \ ,  
\nonumber\\
d_5 &=& \frac{1}{14175} \left[ 
21 + 50 p^2 - 21 p^4 + 10 p^6 - 3 p^8 \right] \ ,
\ {\rm etc.}
\label{djs}
\end{eqnarray}
The results for the induced perturbative (\ref{U0}) and
nonperturbative parts (\ref{Un}), and their ratios, 
are presented in Figs.~\ref{Uvsa} (a)--(b) and
\ref{Uratio}, respectively,
in analogy with Figs.~\ref{Lvsa} (a)--(b) and \ref{Lratio}.
The case of the pure
(quasi)magnetic field is not included 
in Figs.~\ref{Uvsa}--\ref{Uratio}, because in this case 
$\delta {\tilde {\cal U}}\!=\!- \delta {\tilde {\cal L}}$ 
and thus the relevant information is already
contained in Figs.~\ref{Lvsa}--\ref{Lratio}.
The behavior of the induced energy densities is,
in broad qualitative terms, similar to that 
of the induced Lagrangian densities.\footnote{
The total induced energy densities, and
values of the truncated perturbation series (\ref{Upert})
that include terms $\sim\!{\tilde a}^8$, are
included in Figs.~\ref{UPade} in Section IV.}

In the special case of QED,
similarly as for the Lagrangian densities 
in the previous Section, 
the total induced energy densities account
for a very small part of the total
energy density ($0.2$--$0.3$ per mille when 
${\tilde a}\!\approx\!1$) and can become
significant only when the field becomes exceedingly
large (${\tilde a}\!\stackrel{>}{\sim}\!10^2$). 
The same is true also for the dielectric
permeability tensor $\varepsilon_{ij}$: 
In the direction of the fields, we have 
$\delta \varepsilon_{\|}\!\equiv\!\varepsilon_{\|}\!-\!1
= a \partial( {\rm Re} \delta {\cal L} )/\partial a|_b$,
i.e., by (\ref{delUk})--(\ref{notU}) we have
$\delta \varepsilon_{\|}\!=\!(
\delta {\tilde {\cal U}}\!+\!{\rm Re} 
\delta {\tilde {\cal L}}) \alpha/(2 \pi)$, 
which is about $10^{-3}$ for
${\tilde a}\!\approx\!1$ and $p\!=\!1$.
Therefore, the effective coupling parameter along the
field direction $\alpha_{\|}\!=\!\alpha/\varepsilon_{\|}$
changes by about one per mille, while 
$\alpha_{\perp}\!=\!\alpha/\varepsilon_{\perp}$
remains unchanged since $\varepsilon_{\perp}\!=\!1$.
Therefore, in QED, 
any quantity which can be expanded in powers of
the coupling parameter alone (without fields)
remains a perturbative quantity.
QED then remains a perturbative theory
despite such strong fields -- 
cf. also Ref.~\cite{Peccei:1988wh} on that point.

The energy density is not stable in time
when ${\tilde a}\!\not=\!0$, due to the
energy losses to pair creation of fermions of mass $m$. 
It decreases by about 50 percent in the time $t_{1/2}$
\begin{equation}
t_{1/2} \approx \frac{\pi^2}{8 \alpha}
\exp \left( + \frac{\pi}{{\tilde a}} \right)
\left[\frac{ ( 1 + p^2) }{ p \pi \coth( p \pi) }
\right]
\frac{1}{m} \ , 
\label{halftime}
\end{equation}
where $\alpha\!\equiv\!g^2/(4 \pi)$.
The factor in the square brackets, appearing
due to the presence of the (quasi)magnetic
field, is usually not essential in the estimates
since it is $\sim$$1$ for $p\!\leq\!5$.
In the case of QED and with $p\!=\!0$,
$t_{1/2}$ is about $0.9 \cdot 10^5 m_e^{-1}$,
$0.4 \cdot 10^4 m_e^{-1}$ and $0.3 \cdot 10^3 m_e^{-1}$
for ${\tilde a} = 0.5$, $1$, and $5$, respectively.
Here, $m_e^{-1} \approx 1.3 \cdot 10^{-21}$ s
is the electron Compton time.

\section{Quasianalytic continuation into 
the nonperturbative sector}

In this Section, we use the discussed induced
densities as an example on which to test
and get some insights into
methods of approximate analytic (i.e., quasianalytic)
continuation. In various physical contexts,
such methods allow one to extract all
or part of the information on the
nonperturbative sector from the knowledge
of the perturbative sector alone. We will
use the method of Borel--Pad\'e transformation,
or a modification thereof.

One may ask whether the perturbation
expansions (\ref{Lpert}) and (\ref{Upert})
allow us to obtain the full, including the nonperturbative,
information about the corresponding densities.
The answer for the Lagrangian density is yes, 
but under the condition that we take 
in the corresponding Borel--Pad\'e approximants
the Cauchy principal values when integrating
over the positive poles of the Pad\'e integrand
in the inverse Borel transformation. This is
reflected in the terms ${\rm i}{\epsilon}$ in the denominators
of the integrands of (\ref{EH2}) and/or (\ref{Ln}).
More specifically, we first Borel--transform (B)
the perturbation series (\ref{Lpert})
\begin{equation}
B \left[ \frac{\delta {\tilde {\cal L}}^{\rm pert.}({\tilde a};p)}
{\tilde a} \right] = c_1(p) {\tilde a} + c_3(p) {\tilde a}^3
+ c_5(p) {\tilde a}^5 + \cdots  \ ,
\label{BL}
\end{equation}
then construct an $[N/M]_{{\rm B}}({\tilde a};p)$ 
Pad\'e approximant to B of (\ref{BL}),\footnote{
$[N/M]({\tilde a})$ Pad\'e approximant to
(\ref{BL}) is defined by two properties:
1.~it is a ratio of two polynomials
in ${\tilde a}$, the nominator polynomial
having the highest power ${\tilde a}^N$ and
the denominator ${\tilde a}^M$; 2.~when expanded
in powers of ${\tilde a}$, it reproduces the
coefficients at the terms ${\tilde a}^n$ in (\ref{BL})
for $n\!\leq\!N\!+\!M$; it is based solely on the
knowledge of these latter 
coefficients $c_n$ ($n\!\leq\!N\!+\!M$).} 
and then apply the inverse Borel transformation
\begin{equation}
BP^{\rm [N/M]} \left[ \delta {\tilde {\cal L}}^{\rm pert.}
\right]({\tilde a};p) =
\int_0^{\infty}
dw \exp \left( - \frac{w}{{\tilde a}} \right)
[N/M]_{\rm B}(w;p) \ .
\label{BPL}
\end{equation}
On the other hand, the real part of the actual density
(\ref{EH2}) can also be written as a Borel--type integral,
when introducing $w\!\equiv\!a g z$ 
and ${\epsilon}^{\prime}\!\equiv\!a g {\epsilon}$ in (\ref{EH2})
and normalizing the density according to (\ref{not})
\begin{eqnarray}
{\rm Re} \delta {\tilde {\cal L}} &=& 
{\rm Re} \int_0^{\infty}
d w \exp \left(- \frac{w}{\tilde a} \right)
\frac{(-1)}{w}
\left[ \frac{p \cos(w)}{\sin(w\!+\!{\rm i}{\epsilon}^{\prime})} 
\coth(p w) + \frac{1}{3} (1\!-\!p^2) - \frac{1}{w^2} \right] \ .
\label{EH4}
\end{eqnarray}
The expansion of the integrand of (\ref{EH4}),
excluding the exponential, in powers of $w$
is identical with the Borel transform (\ref{BL})
with ${\tilde a}\!\mapsto\!w$, as it should be. 
Comparing (\ref{BPL}) with the exact result (\ref{EH4}),
we see that the Borel--Pad\'e method (\ref{BPL}) will
be efficient in (quasi)analytic continuation if
Pad\'e approximants
$[N/M]_{{\rm B}}(w)$ approach the 
integrand of (\ref{EH4}) in an increasingly wide integration
interval of $w$ when the Pad\'e order indices $N$ 
and $M$ ($\approx\!N$) increase.
This in fact happens, since the integrand
in (\ref{EH4}) is a meromorphic function
in the complex plane whose poles structure
on the positive axis is especially
simple -- there are only single (not multiple) 
poles, located at $w\!=\!\pi, 2 \pi, 3 \pi$.
Pad\'e approximants to power expansions of such functions 
are known to approximate such functions 
increasingly better when the Pad\'e order
indices $N\!\approx\!M$ increase \cite{Baker}.
Near the poles $w\!\approx\!n \pi$ the
integrand behaves as
$\sim$$(w\!-\!n \pi\!+\!{\rm i}{\epsilon}^{\prime})^{-1}$.
Hence, for obtaining the real (dispersive) part
of the density, the Borel integration over the poles
has to be taken with the Cauchy principal value (CPV)
prescription -- not just in the exact expression
(\ref{EH4}), but also in the approximate expression
(\ref{BPL}). Thus, the Borel integration
in (\ref{BPL}) over the $n$'th pole, i.e., the $n$'th
renormalon contribution, has in the case
at hand no renormalon ambiguity. As the Pad\'e
order indices $N\!\approx\!M$ are increased,
we thus systematically approach the exact
${\rm Re} \delta {\tilde {\cal L}}$
via the CPV of (\ref{BPL}).
This means that in the case at hand [strong
(quasi--)Abelian fields with fermionic fluctuations
included], the full induced Lagrangian density can
be obtained on the basis of the knowledge of perturbation 
expansion (\ref{Lpert}) for weak fields and the
CPV prescription. The more
terms in (\ref{Lpert}) [and thus in (\ref{BL})]
we know, the higher Pad\'e order indices $N\!\approx\!M$ 
we can have, and hence the closer to the full 
Lagrangian density we can come via (\ref{BPL}).

\begin{figure}
\noindent
\begin{minipage}[t]{.49\linewidth}
 \centering\epsfig{file=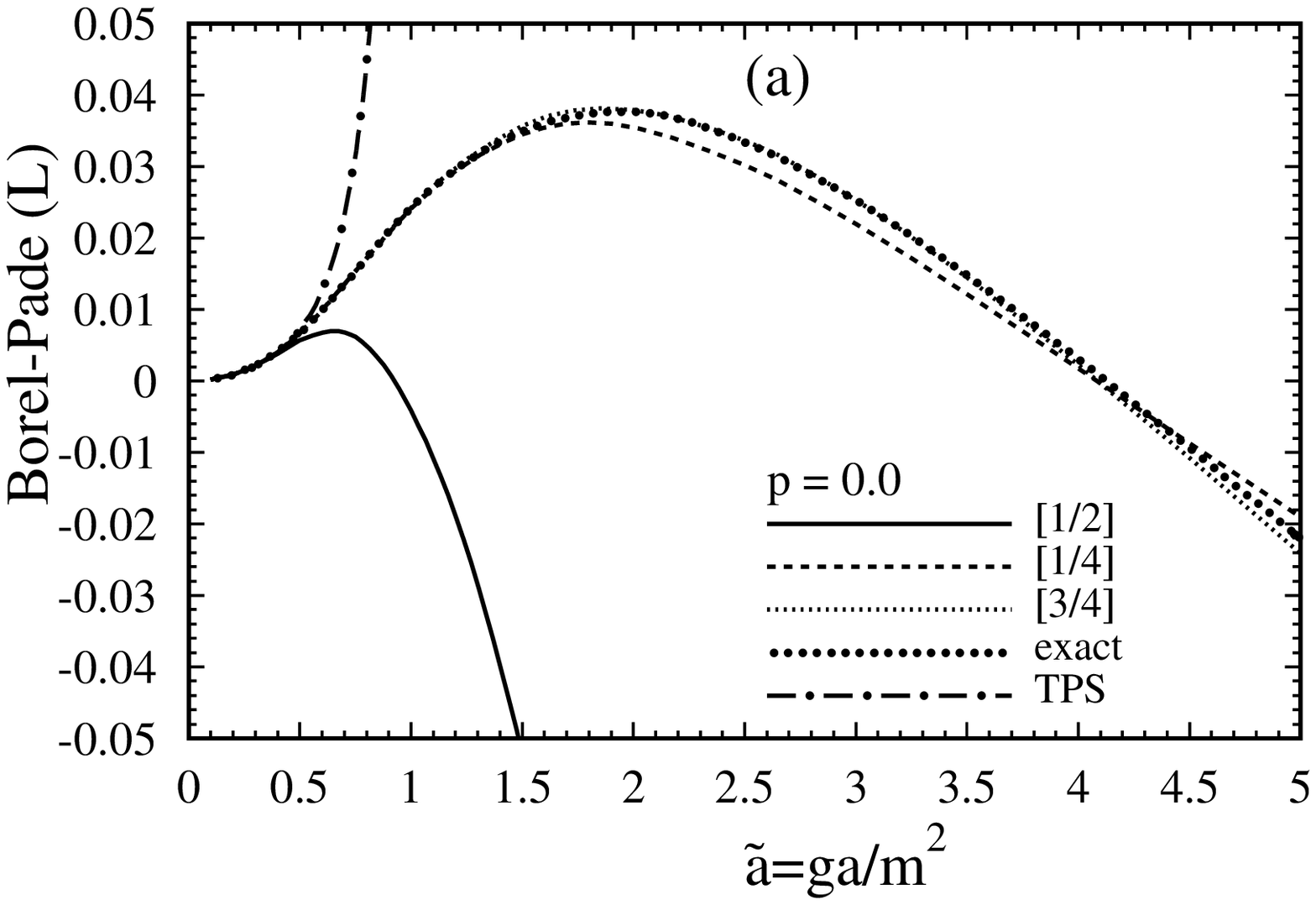,width=\linewidth}
 \centering\epsfig{file=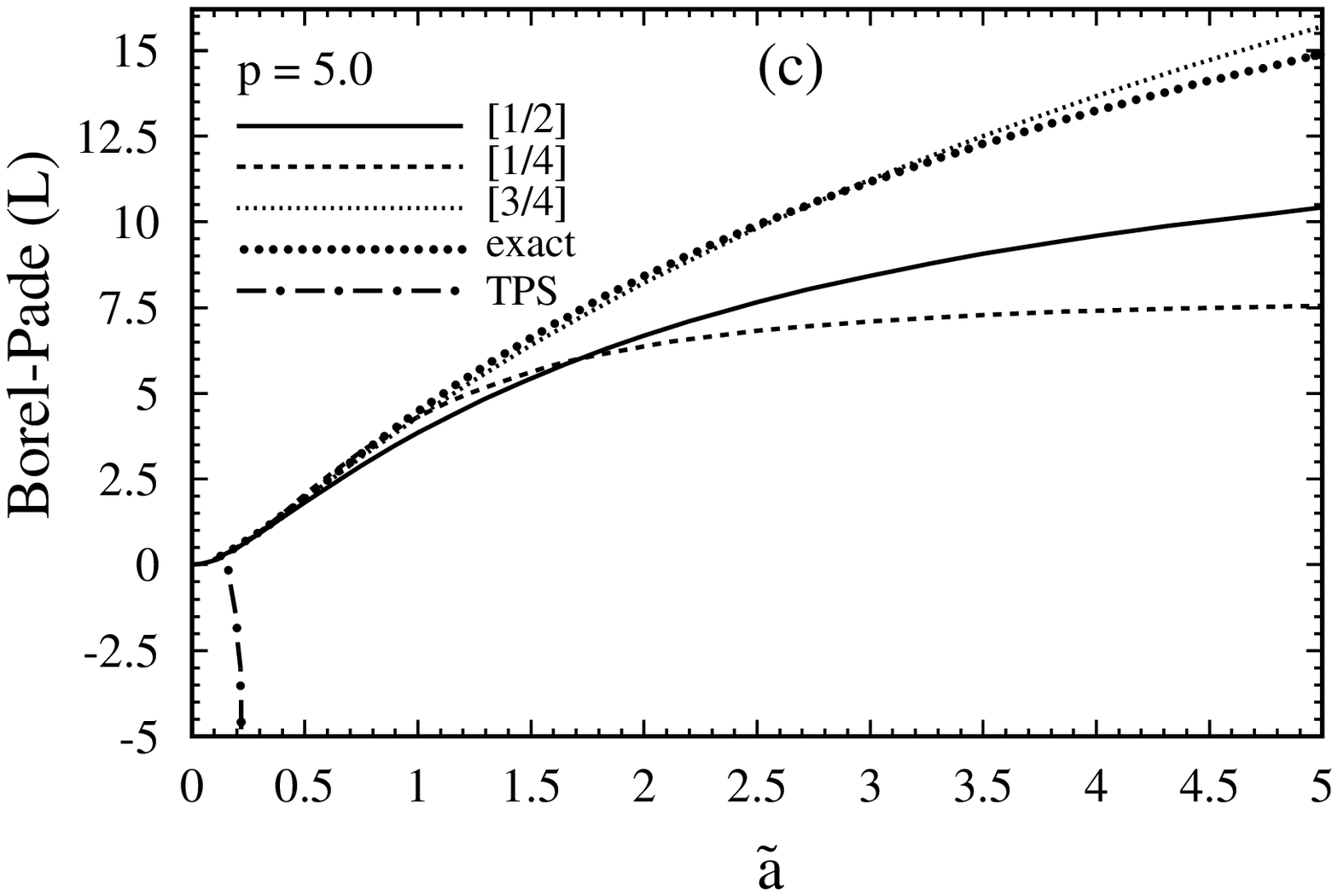,width=\linewidth}
\end{minipage}
\begin{minipage}[t]{.49\linewidth}
 \centering\epsfig{file=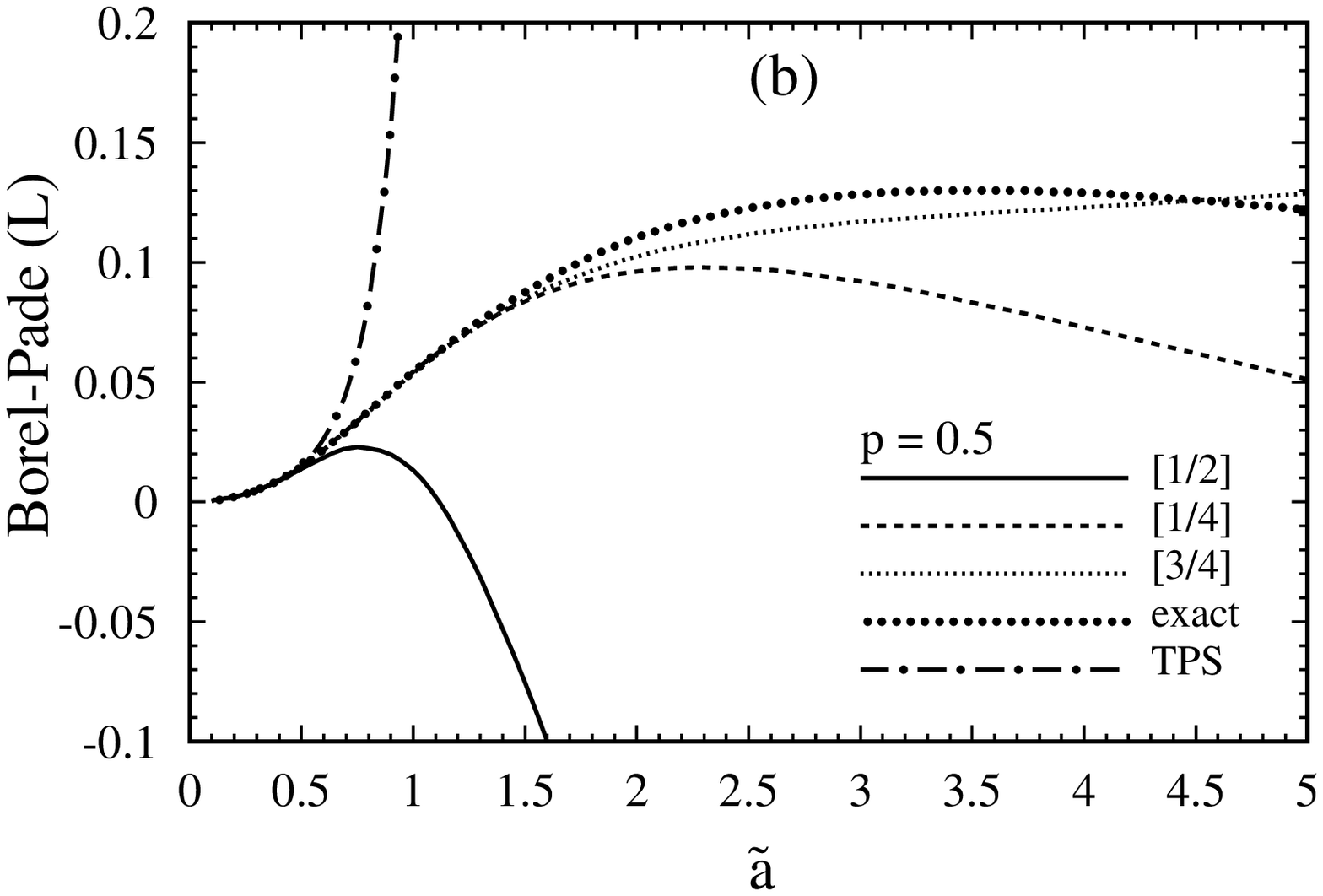,width=\linewidth}
 \centering\epsfig{file=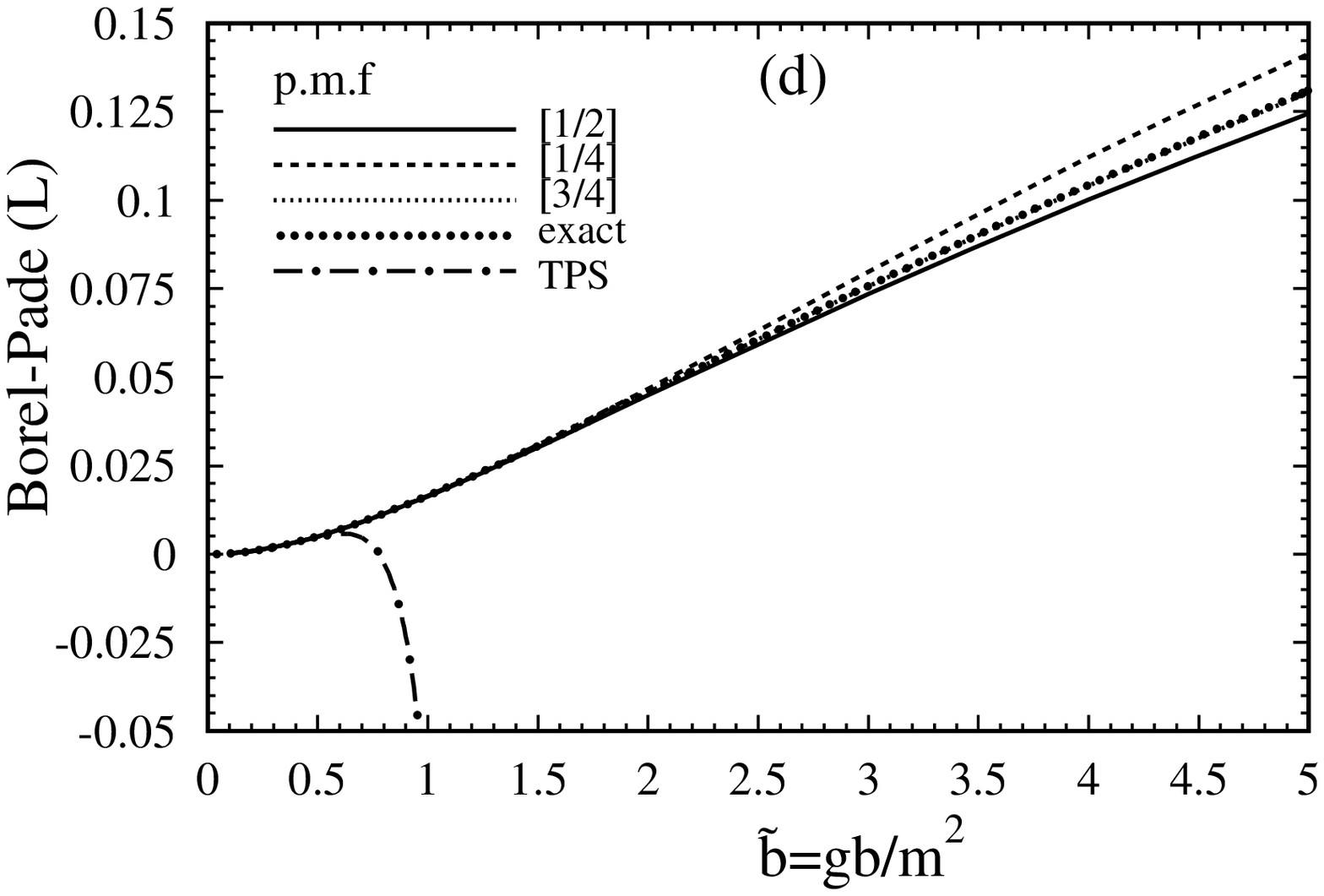,width=\linewidth}
\end{minipage}
\caption{\footnotesize
Borel--Pad\'e approximants (BP's) to the induced 
dispersive Lagrangian density (\ref{not}) as functions
of ${\tilde a}$, for various values of
$p\!=\!{\tilde b}/{\tilde a}$: (a) $p\!=\!0$;
(b) $p\!=\!0.5$; (c) $p\!=\!5.0$; (d) pure magnetic field 
(${\tilde a}\!=\!0$).
Depicted are those BP's (\ref{BPL}) which are
based on the Pad\'e approximants $[1/2]$,
$[1/4]$ and $[3/4]$ of the Borel--transform (\ref{BL}).
The numerically exact curves 
[sum of curves of Figs.~\ref{Lvsa} (a) and (b)] are
also included and they virtually agree with the
$[3/4]$ BP results.
Included are also the the results of the truncated perturbative
series which include terms $\sim$${\tilde a}^8$
[in (d): $\sim$${\tilde b}^8$].} 
\label{LPade}
\end{figure}

On the basis of the knowledge of the first four 
nonzero perturbation terms in (\ref{Lpert}) and
correspondingly in (\ref{BL}), we can construct
the following Pad\'e approximants of the
perturbative Borel transform (\ref{BL}):
$[1/2]_{{\rm B}}, [1/4]_{{\rm B}}, [3/4]_{{\rm B}}$.
Then we can calculate the corresponding
Borel-Pad\'e transforms via (\ref{BPL}) with the CPV
prescription. The corresponding results
of the approximants for the full induced density
${\rm Re} \delta {\tilde {\cal L}}$ are presented
in Fig.~\ref{LPade}, together with the exact
numerical values calculated by (\ref{EH4}) in Section II. 
The curves are given as functions of the (quasi)electric
strength parameter ${\tilde a}$ for four fixed values of the 
magnetic--to--electric field ratio $p$,
and Fig.~\ref{LPade} (d) is for the case
of the pure (quasi)magnetic field (${\tilde a}\!=\!0$). 
We see that the highest order ($[3/4]$) 
Pad\'e--Borel results agree well with the exact results
over the entire depicted region of $\tilde a$.
When the Pad\'e order
indices $N$ and $M$ ($\sim$$N$) increase,
the region of agreement includes increasingly
large values of ${\tilde a}$.
For comparison, we also included the results
of the truncated perturbation series (TPS)
made up of the first four nonzero terms of
(\ref{Lpert}) [in Fig.~(d): for the corresponding p.m.f. case],
i.e., those perturbation terms which the
presented Borel--Pad\'e transforms are based on.

If we apply the very same procedure in the
case of the energy density -- Borel--transforming the
series $\delta {\tilde {\cal U}}^{\rm pert.}/{\tilde a}$
of (\ref{Upert}), constructing Pad\'e approximants,
and carrying out the inverse Borel transformation
by using the Cauchy principal value (CPV) prescription
-- the results are disappointing. It turns out that
increasing the Pad\'e order indices $N$ and $M$ ($\sim$$N$)
does not generally result in a better precision.
For example, for $p\!\stackrel{<}{\approx}0.5$
and ${\tilde a}\!\stackrel{>}{\approx}\!0.5$,
the Borel--Pad\'e transforms of the order
$[3/4]$ and $[3/6]$ give significantly worse
results than those of the lower order $[1/4]$.
The reason for this erratic behavior of the
Borel--Pad\'e approximants in this case lies in
the more complicated poles structure of the Borel--Pad\'e
transforms.
This can be seen if we rewrite $\delta {\tilde {\cal U}}$
in the Borel--integral form analogous to (\ref{EH4}),
obtained from (\ref{EH4}) by applying relation
(\ref{delUk})
\begin{eqnarray}
{\rm Re} \delta {\tilde {\cal U}} &=&
{\rm Re} \int_0^{\infty}
d w \exp \left(- \frac{w}{\tilde a} \right)
\frac{(-1)}{w}
\left[ - \frac{ p w }{\sin^2(w\!+\!{\rm i}{\epsilon}^{\prime})} 
\coth(p w) + \frac{1}{3} (1\!+\!p^2) + \frac{1}{w^2} \right] \ .
\label{U4}
\end{eqnarray}
The expansion in powers of $w$ of the integrand in (\ref{U4}), 
excluding the exponential,
gives of course the exact Borel transform of the perturbation 
series (\ref{Upert}) divided by ${\tilde a}$ (and replacing 
${\tilde a}\!\mapsto\!w$)
\begin{eqnarray}
\lefteqn{
- \frac{1}{w}
\left[ - \frac{ p w }{\sin^2(w\!+\!{\rm i}{\epsilon}^{\prime})} 
\coth(p w) + \frac{1}{3} (1\!-\!p^2) + \frac{1}{w^2} \right]  }
\nonumber\\
&& =  d_1(p) w + d_3(p) w^3 + d_5(p) w^5 + \cdots \equiv
B \left[ \frac{\delta {\tilde {\cal U}}^{\rm pert.}(w;p)}{w}
\right] \ .
\label{BU}
\end{eqnarray}
However, we now see that this integrand has
a double poles structure on the positive
$w$ axis, the double poles located at $w\!=\!\pi,
2 \pi, 3 \pi, \ldots$. The Pad\'e approximants to 
the power series (\ref{BU}) have great trouble
simulating this double poles structure adequately.
When they do it by creating one single 
or two nearby real poles,
say near $w\!=\!\pi$, then it turns out that the
inverse Borel transformation via the Cauchy principal value 
(CPV) prescription often gives good results. However,
when the Pad\'e approximants try to simulate the
double pole near $w\!=\!\pi$ by creating two
mutually complex--conjugate poles $a\!\pm{\rm i} b$
($a\!\approx\!\pi$, $|b|\!\ll\!\pi$),
the inverse Borel transformation 
gives very unsatisfactory results. 
This occurs, for example, in Pad\'e approximants
$[3/4](w;p)$ and $[3/6](w;p)$ for $p\!\leq\!0.5$.
Heuristically we can understand that such a simulation 
is bad, because the structure of the
integrand in (\ref{U4}) suggests that a double pole
at $a\!-\!{\rm i} b$ alone, just below the real axis,
would do a better job, but it is not allowed in the
Pad\'e approximants. The latter is true because
the perturbation expansion (\ref{BU}) is explicitly real
for real $w$'s, and this property is hence shared also by 
the Pad\'e approximants, enforcing for each complex pole
another pole which is complex--conjugate.

To overcome this problem, the idea is to modify the
Borel transformation of the perturbation series
(\ref{Upert}) in such a way that the resulting transformed
series is represented by a (meromorphic) function
without any double poles on the real positive axis,
in contrast to the Borel--transformed series (\ref{BU}).
This, in fact, can be implemented in the easiest way
by using the following modification of the Borel
transformation (MB):
\begin{eqnarray}
\frac{ \partial MB \left[ \delta {\tilde {\cal U}}^{\rm pert.}
\right](w;p) }{ \partial w }
&= & B \left[ \frac{\delta {\tilde {\cal U}}^{\rm pert.}(w;p)}{w}
\right]
= d_1(p) w + d_3(p) w^3 + d_5(p) w^5 + \cdots  \quad
\nonumber\\
\Rightarrow \quad
MB \left[ \delta {\tilde {\cal U}}^{\rm pert.}
\right](w;p) & = & 
d_1(p) \frac{w^2}{2} + d_3(p) \frac{w^4}{4} + 
d_5(p) \frac{w^6}{6} + \cdots \ .
\label{MB}
\end{eqnarray}
This trick changes every double pole in the B into the
corresponding single pole in the MB. Then we apply
Pad\'e approximants $[N/M]_{{\rm MB}}(w)$ to the
MB series (\ref{MB}), and carry out the corresponding
inverse modified Borel transformation
\begin{equation}
MBP^{\rm [N/M]} \left[ \delta {\tilde {\cal U}}^{\rm pert.}
\right]({\tilde a};p) =
\frac{1}{{\tilde a}} \int_0^{\infty}
dw \exp \left( - \frac{w}{{\tilde a}} \right)
[N/M]_{\rm MB}(w;p) 
\label{MBPU}
\end{equation}
with the Cauchy principal value (CPV) prescription
when integrating over the single poles.
This CPV prescription originates again from the
${\rm i} {\epsilon}^{\prime}$ terms in the
double poles structure
$(w\!-\!n \pi\!+\!{\rm i}{\epsilon}^{\prime})^{-2}$
of the Borel--transform (B) integrand of (\ref{U4})
that is now changed to the single poles structure
$(w\!-\!n \pi\!+\!{\rm i}{\epsilon}^{\prime})^{-1}$
in the modified Borel--transform (MB) of (\ref{MB})
whose Pad\'e approximants $[N/M]_{\rm MB}(w;p)$
appear in (\ref{MBPU}).
  
\begin{figure}
\begin{minipage}[t]{.49\linewidth}
 \centering\epsfig{file=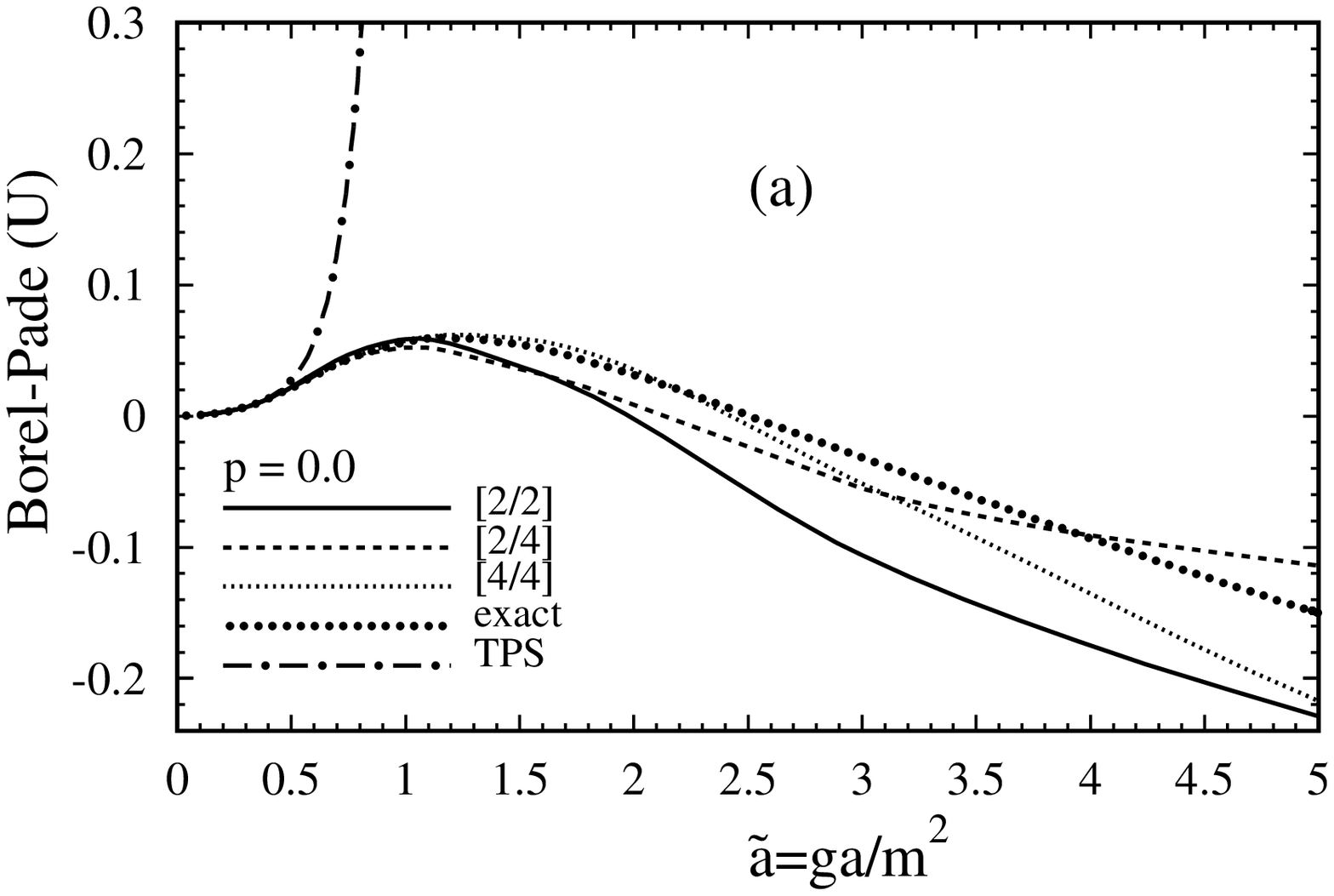,width=\linewidth}
 \centering\epsfig{file=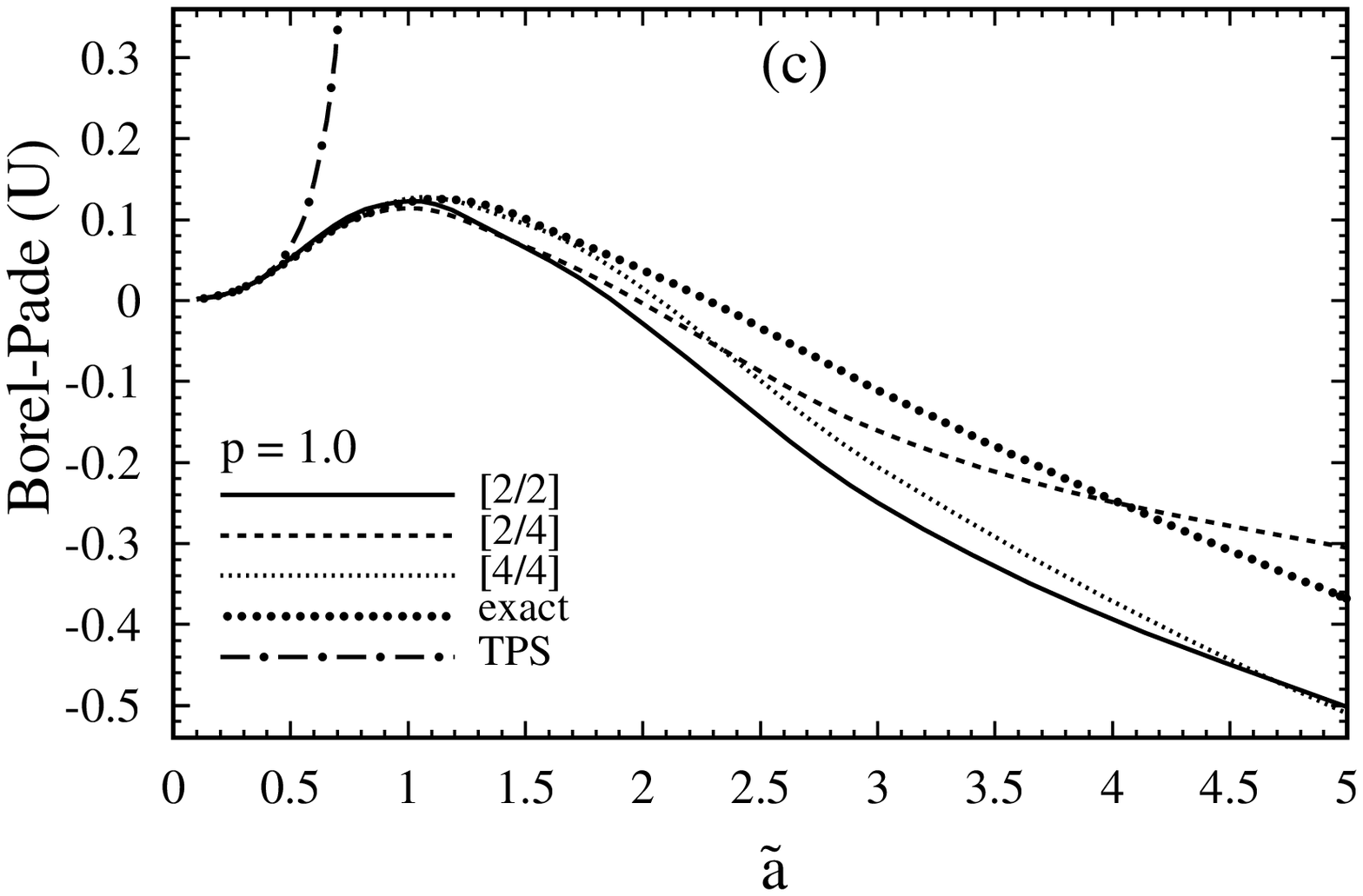,width=\linewidth}
\end{minipage}
\begin{minipage}[t]{.49\linewidth}
 \centering\epsfig{file=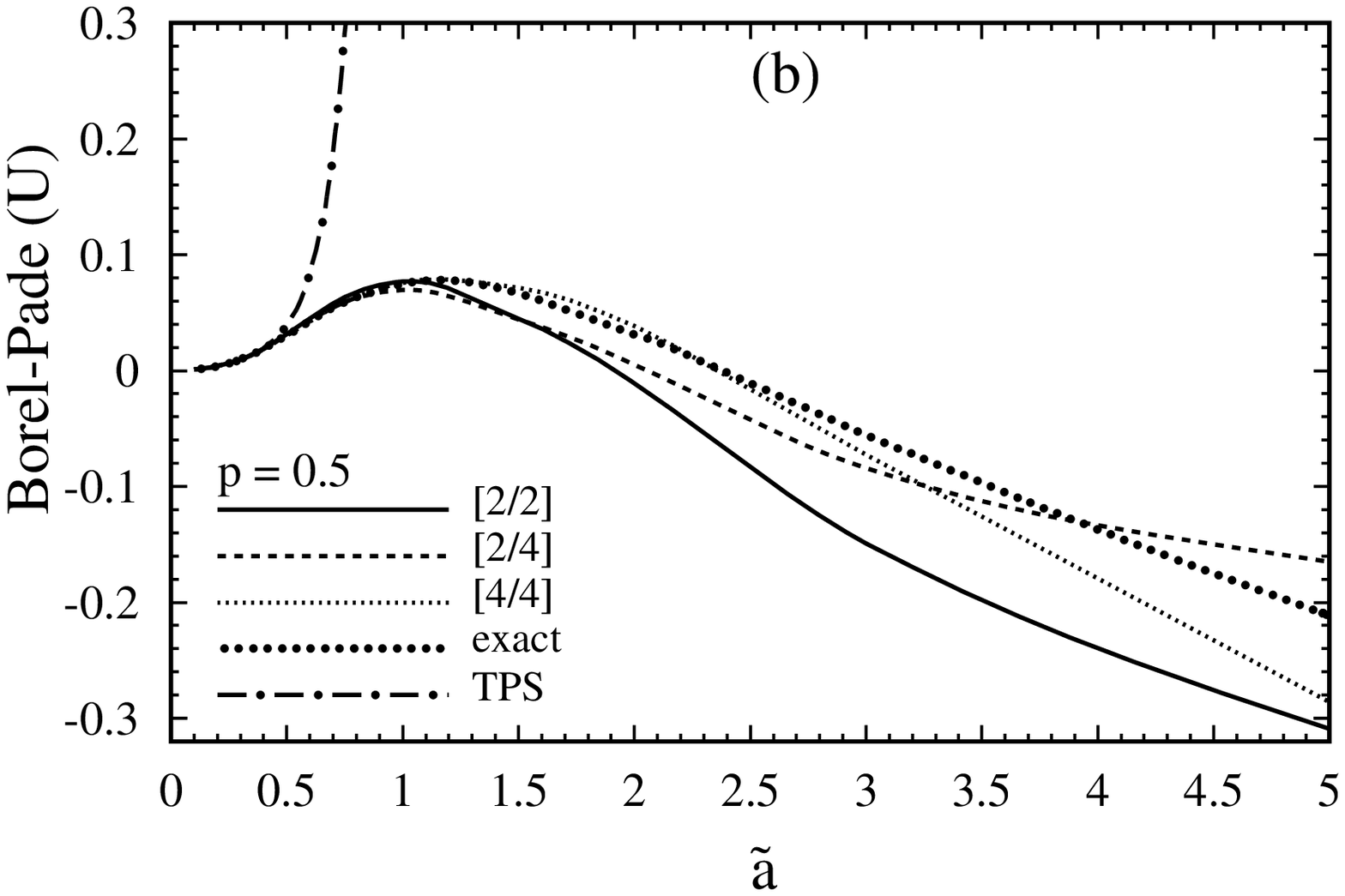,width=\linewidth}
 \centering\epsfig{file=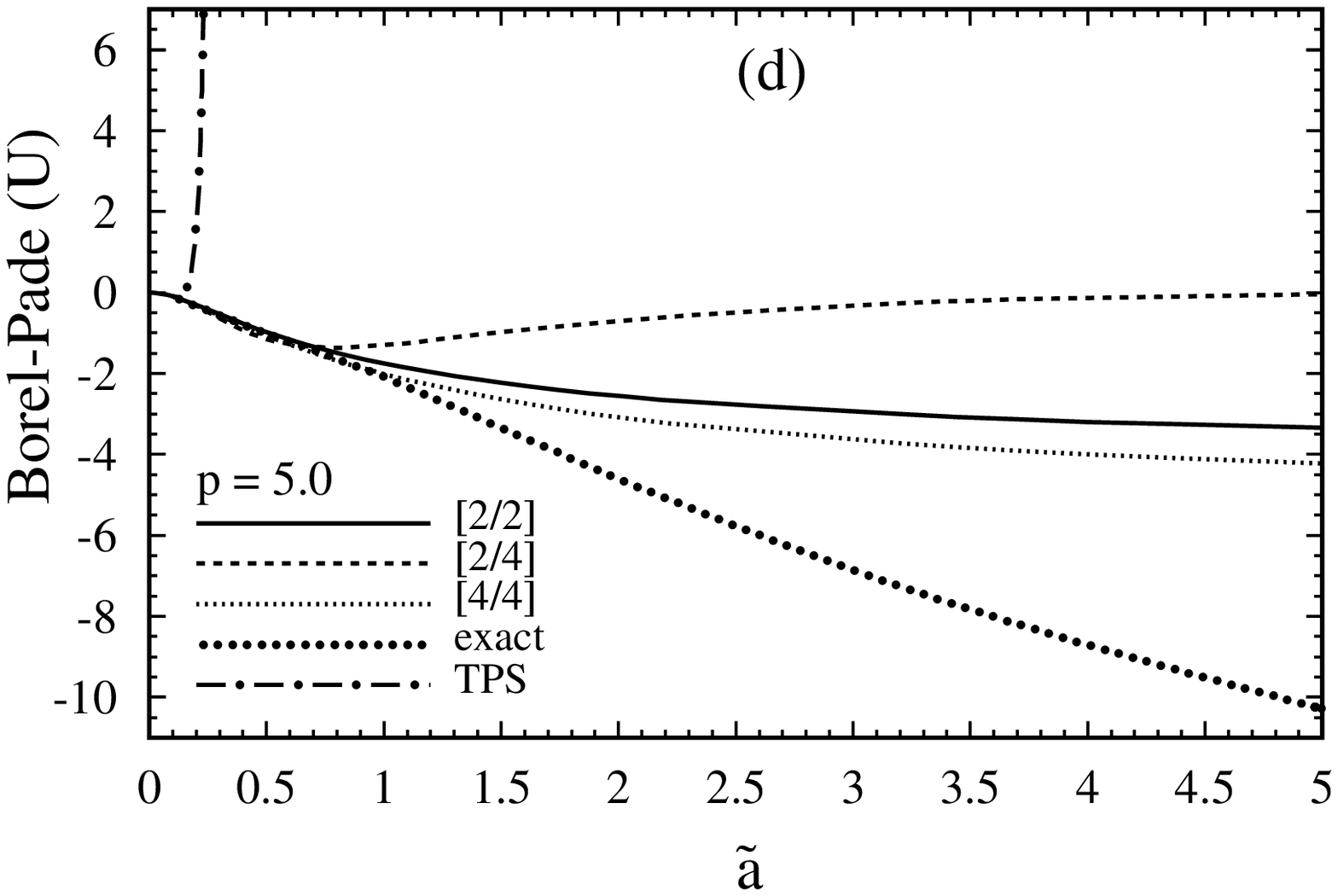,width=\linewidth}
\end{minipage}
\vspace{0.0cm}
\caption{\footnotesize
Modified Borel--Pad\'e approximants [MBP's -- cf.~(\ref{MBPU})] 
to the induced energy densities, based on the Pad\'e
approximants $[2/2]$, $[2/4]$ and $[4/4]$ for the MBP's (\ref{MB}),
as functions of ${\tilde a}$, at fixed
values of $p\!=\!{\tilde b}/{\tilde a}$:
(a) $p\!=\!0$; (b) $p\!=\!0.5$; (c) $p\!=\!1.0$; (d) $p\!=\!5.0$.}
\label{UPade}
\end{figure}

The numerics clearly confirm that these MBP's 
(\ref{MBPU}) are well behaved, i.e., 
they approximate well the actual
full induced energy density 
$\delta {\tilde {\cal U}}({\tilde a};p)$
in the region of ${\tilde a}$ 
which is getting wider when the
Pad\'e order indices $N$ and $M$ ($\approx\!N$)
increase. The results are presented in 
Fig.~\ref{UPade}, where the MBP's for the
first three possible Pad\'e order indices
$[2/2]$, $[2/4]$ and $[4/4]$, along with the
exact numerical results, are shown as functions
of ${\tilde a}$, at four fixed values of 
$p\!\equiv\!{\tilde b}/{\tilde a}$.
Another reason why the results now behave better
than those of the usual BP tranforms lies in the
fact that the Pad\'e approximants
($[2/2]$, $[2/4]$ and $[4/4]$) are now more diagonal
than earlier ($[1/2]$, $[1/4]$ and $[3/4]$).
This is due to one additional power of $w$ in the MB
series (\ref{MB}), as compared with the 
usual B series.
The diagonal and near--diagonal Pad\'e approximants 
are known to behave better than the (far) 
off--diagonal ones \cite{Baker}.
In fact, Figs.~\ref{UPade} suggest that clear
improvement -- extension of the ${\tilde a}$ range
of agreement with the exact results -- 
sets in when we switch from $[2/2]$
to $[4/4]$ MBP, while the off--diagonal
$[2/4]$ MBP may even be slightly worse
than $[2/2]$.
For comparison, we also included the results
of the truncated perturbation series (TPS)
made up of the first four nonzero terms 
(up to $\sim$${\tilde a}^8$) of (\ref{Upert}),
i.e., the terms on which the presented Borel--Pad\'e 
transforms are based. The case of the pure
(quasi)magnetic field was not included in
these Figures because in this case
$\delta {\tilde {\cal U}}\!=\!-\delta {\tilde {\cal L}}$
and thus the information on this case is
contained in Fig.~\ref{LPade} (d).

This application of Borel--Pad\'e transformations
and their modification may give us some insights
into how the (quasi)analytic continuation from
the perturbative (small ${\tilde a}$) into the
nonperturbative (large ${\tilde a}$) regions
can be carried out in other theories whose
exact behavior in the latter region is still theoretically
unknown. One such example is the perturbative QCD (pQCD),
where some observables are known at the 
next--to--next--to--leading order (${\rm N}^2{\rm LO}$).
The coupling parameter in that case 
[${\tilde a} \mapsto \alpha_s(Q^2)$] 
can be quite large when the relevant
energies of the process are low ($Q\!\sim\!1$ GeV),
thus rendering the direct evaluation of the 
${\rm N}^2{\rm LO}$ TPS
unreasonable or at best unreliable. 
When applying Borel--Pad\'e transformations or
modifications thereof to such series, we are faced
with two major problems:
\begin{itemize}
\item
The first problem is of a more technical nature.
Since only very few, at most two, coefficients
beyond the leading order are known, the Pad\'e
approximants associated with the (modified) Borel
transform of the series have low order indices
($N, M \leq 2$) and thus do not necessarily
reproduce the location of the leading poles on the 
positive axis, if they exist, adequately.   
\item
The second problem is of a deeper theoretical nature.
Knowing too little about the behavior of QCD in,
or close to, the nonperturbative regime,
we do not know how to integrate
over the possible positive poles in the inverse
(modified) Borel transformation -- this can be
termed the infrared renormalon ambiguity \cite{renormalon}.
\end{itemize}

In the discussed case of integrated fermionic fluctuations 
in strong (quasi--)Abelian fields -- for the Lagrangian 
and energy densities -- we do not face any of the
two afore--mentioned problems since the exact
solution is known. We have to
apply the Cauchy principal value (CPV) prescription in the
integration of the Borel--Pad\'e transform of the
induced dispersive Lagrangian density, and in 
the modified Borel--Pad\'e transform of the 
induced energy density. 
The CPV is the direct consequence of the path
(${\epsilon}$ parameter) in the exact solution
(\ref{EH1}) [$\Leftrightarrow$ (\ref{EH2})].
The knowledge
of the full theoretical solution in the latter
case also tells us that the poles structure
of the usual Borel transform of the induced energy
density is more complicated (double poles),
so that we have to apply a modified
Borel-Pad\'e transform which changes the
double poles into a single poles structure.

We point out that the positive poles -- renormalons --
discussed in the present work cannot be
directly identified with the usual infrared 
(ultraviolet) renormalons in QCD (QED).
The latter renormalons,
as defined in the literature \cite{renormalon},
are interpreted in the perturbative language
as originating from renormalon chains at 
low (high) momenta $k$.
The renormalon chains are momentum--$k$
gluon (photon) propagators with $n$ chained
one--loop insertions, where $n$ can be
arbitrarily large. In the model at hand, however,
only quantum fluctuations of fermions, 
in the slowly--varying strong fields, are considered;
the effects of the quantum fluctuations of propagating gluons 
(photons) were not included in the discussed effective
model. The positive poles, i.e. renormalons,
in the present model originate from a collective effect
of arbitrarily many very soft gluons (photons)
coupling to a fermion loop or to a fermion propagator
-- cf.~\cite{ChiuNussinov}. The relevant parameter of 
the effective coupling 
of these soft gauge bosons to the fermions,
appearing in the induced effective action,
is ${\tilde a}\!=\!g a/m^2$
and it can be large due to the strong field $a$
and/or due to the strong coupling $g$.
These nonperturbative contributions are then
roughly $\sim$$\exp(- {\rm const.\/}/{\tilde a})
\!=\!\exp[- {\rm const.} m^2/(g a)]$
-- cf.~(\ref{BPL}), (\ref{MBPU}).
This is similar, but not identical, 
to the infrared renormalon
contributions in QCD 
$\sim$$\exp(- {\rm const.^{\prime}\/}/g^2)$.
We may be tempted to term the
renormalons discussed in the present
paper as infrared renormalons due
to their nonperturbative origin in the infrared,
although this name is reserved for the
afore--mentioned QCD--type renormalons.

Various QCD and QED applications of the
Borel-Pad\'e approach, with CPV prescription,
have been made in \cite{Raczka}--\cite{Jentsch}.
The new method of Ref.~\cite{Jentsch}
gives modified real and imaginary
parts of the Borel--Pad\'e of $\delta {\tilde {\cal  L}}$,
in comparison to the usual CPV prescription,
when the Pad\'e approximants $[N/M]_B$
have poles off the positive real axis.
This may influence the speed of the convergence
of the Borel--Pad\'e transforms towards the
full solution when the Pad\'e order indices
$N$ and $M$ ($\approx\!N$) increase.
This method came to our attention after
finishing the manuscript.

Two other references \cite{Dunne:1999uy}--\cite{Jentschura:1999vn}
are also somewhat related to our work.
Dunne and Hall \cite{Dunne:1999uy}
considered, among other things, the question of
resummation of the (one--loop) Euler--Heisenberg (EH)
Lagrangian density by using the knowledge of the
perturbation expansion of the Borel transform.
Since they did not use Pad\'e in addition, they
needed at least an approximate information on {\em all\/}
the coefficients of the series to reconstruct
approximately the nonperturbative sector.
Jentschura {\em et al.\/} \cite{Jentschura:1999vn},
on the other hand, did not employ the Borel transform,
but applied directly to truncated perturbation series (TPS)
of the EH density a numerical method
(Weniger sequence transformation)
which differs from Pad\'e in several aspects.
Their results of resummation are better than the direct
application of Pad\'e to the TPS of the EH Lagrangian,
but they are worse than the results of the
combined Borel--Pad\'e method.

\section{Conclusions}

We introduced the concept of separation
of the induced dispersive action
into the nonperturbative and perturbative parts.
We then investigated numerically
the nonperturbative contributions to the
dispersive (real) part of the Lagrangian density
and to the real energy density, induced
by quantum fluctuations of fermions in
the strong (quasi--)Abelian fields that
don't change significantly in space--time over the
typical fermionic Compton wavelengths $1/m$.
There are only nonperturbative contributions
in the absorptive (imaginary) part
of the strong field Lagrangian density, 
the latter part being responsible for the 
fermion--antifermion pair creation.
On the other hand, 
the nonperturbative contributions in the real
(dispersive) sector are in general also significant
and can often even dominate over the perturbative
induced contributions there. 
The induced dispersive
Lagrangian density modifies the Maxwell equations
for strong fields. The induced energy density is
in principle an observable quantity.
When the (quasi)electric fields are strong,
however, these densities decay fast
(in $\sim$$10^3$ Compton times, for ${\tilde a}\!\sim\!1$).
These two induced densities lead to a change in the
dielectric permeability tensor of the vacuum. In the special case
of QED, all these induced effects are below one per cent 
unless the fields are huge (${\tilde a}\!\sim\!10^2$).

We then used the discussed induced quantities
as a ``laboratory'' to test and investigate the efficiency
of specific methods of quasianalytic continuation
from the perturbative region (weak fields) 
into the nonperturbative region (strong fields).
We employed the method of Borel--Pad\'e for the induced 
dispersive Lagrangian density, since the function
represented by the Borel transform series has only simple poles.
For the induced energy density, we had to
employ a modified Borel--Pad\'e transformation
since the function represented by the (nonmodified) 
Borel transform series has double poles.
We found out numerically that such quasianalytic 
continuations become precise over an
increasing region of the effective expansion
parameter ${\tilde a}$ when the number of available
terms in the perturbative expansion increases. 
This means that the quasianalytic continuation gradually 
becomes the analytic (exact) continuation when the
number of the perturbative expansion terms
accounted for increases. The Borel integration over 
positive poles (renormalons) is necessary.
The correct prescription for the integration over
these poles, in the case at hand, is the simplest one -- 
the Cauchy principal value (CPV) prescription,
its origin being the path
(${\epsilon}$ parameter) in the exact solution
(\ref{EH1}) [$\Leftrightarrow$ (\ref{EH2})].
Such analyses could give us some insight into the
problems faced in QCD when nonperturbative 
contributions to observables are investigated
either on the basis of the perturbative results
themselves or by using other models \cite{Benekeetal}
that are at least partly motivated by perturbative methods.

The correct analytic continuation,
in the discussed case of strong background gauge fields,
is the one employing the simplest (CPV) prescription
for integration over the poles in the Laplace--Borel integral.
This appears to be in agreement with the 
conclusions of Ref.~\cite{Matinian:1978mp}
which were obtained from quite different 
considerations involving the renormalization group
-- that the vacuum polarization induced by
the intense gauge fields is in principle determined
by the information on the behavior of the theory
in the perturbative region. The situation in QCD
is less clear. A necessary condition for the existence of 
the (correct) analytic continuation from the perturbative 
into the nonperturbative regime in QCD is that
a nontrivial infrared stable fixed point exist for the 
running strong coupling parameter. 
Such an infrared stable fixed point, however, 
seems to exist only if the number of the quark flavors
is high ($N_f > 9$) \cite{Gardi:1998qr}. For the real
(low--$N_f$) QCD, a phase transition takes place,
and methods of analytic continuation have probably
only a limited range of applicability. 
Stated differently, in this case
the full knowledge of the perturbative sector
probably does not allow us to obtain information on
the deep nonperturbative sector. In such a case,
it is probable that even the renormalon
ambiguity in the low--flavor perturbative QCD (pQCD)
is an intrinsic ambiguity that cannot be entirely eliminated 
with pQCD--related methods alone.

\acknowledgments

The work of G.C. was supported by the Korean
Science and Engineering Foundation (KOSEF).
The work of J.-Y.Y. was supported by the
German Federal Ministry of Science (BMBF).

\begin{appendix}

\section[]{On the analyticity of $\delta {\tilde {\cal L}}_0$ }
\setcounter{equation}{0}

In this Appendix we will clarify the nature of the nonanalytic 
terms $\sim$$\exp(-{\rm const.\/}/{\tilde a})$ that appear
in the naive expansion of the perturbative part
${\rm Re} \delta {\tilde {\cal L}}_0 ({\tilde a};p)$ 
of (\ref{L0}) around the point ${\tilde a}\!=\!0$.
Such terms may in principle be dangerous for our interpretation
of (\ref{L0}) as the perturbative part of the induced
Lagrangian density, because they have the nonanalytic structure
similar to those terms that appear in the nonperturbative parts
${\rm Re} \delta {\tilde {\cal L}}_n ({\tilde a};p)$
of (\ref{Ln}), the latter containing genuinely
nonperturbative contributions due to the singular
(pole) structure of the integrand. We will show that
the mentioned terms in (\ref{L0}) are an artifact of
having the abrupt infrared (IR) cutoff there, and that 
they disappear as soon as the abruptness of the infrared
cutoff is (infinitesimally) softened. 
 
In the proper--time formalism, the IR and UV regions
correspond to the high and the low values of the proper time,
respectively \cite{Ball}.
In the proper--time integral (\ref{EH2})
for $\delta {\cal L}$, the IR region of large
proper time $z$ [$z \geq \pi/(a g)$]
contains poles, the latter leading to
nonperturbative effects. The region of
smaller $z$ has no such singularities and thus
no nonperturbative effects. Therefore, the
perturbative part of $\delta {\cal L}$
should cover the latter region,
and suppress the IR region. The general
way to do this is to introduce,
in the spirit of approaches of \cite{Ball}, 
a nonnegative regulator 
$\rho_{\varepsilon}(w)$ ($w\!\equiv\!a g z$)
in the proper--time integral
\begin{eqnarray}
{\rm Re} \delta {\tilde {\cal L}}^{({\rm P.\/})}_{\varepsilon} 
&=& - {\rm Re} \int_0^{\infty}
\frac{d w}{w} \rho_{\varepsilon}(w)
\exp \left(- \frac{w}{\tilde a} \right) 
\left[ p \cot (w + {\rm i} {\epsilon}^{\prime}) \coth(p w)
+ \frac{1}{3} (1\!-\!p^2) - \frac{1}{w^2} \right] \ ,
\label{Lreg1}
\end{eqnarray}
where the minimal IR regularization requirements are
\begin{equation}
\rho_{\varepsilon}(w) \approx 1 \quad {\rm for} \ w \ll 1 \ ,
\qquad
\rho_{\varepsilon}(w) \ll 1 \quad {\rm for} \ 
w \stackrel{>}{\approx} \pi \ .
\label{reg1}
\end{equation}
The nonnegative parameter ${\varepsilon}$ indicates that we can
choose a class of such regulators. In fact, we will require that
for small $\varepsilon$ a large chunk of the perturbative region, 
namely the $w$--region of approximately $[0,\pi/2]$,
survive in (\ref{Lreg1}). Thus we restrict the minimal conditions
(\ref{reg1}) to the following ones, when ${\varepsilon} \ll 1$: 
\begin{equation}
\rho_{\varepsilon}(w) \approx 1 \quad {\rm for} \ 
w \stackrel{<}{\approx} \pi/2 - \sqrt{\varepsilon} \ ,
\qquad
\rho_{\varepsilon}(w) \ll 1 \quad {\rm for} \ 
w \stackrel{>}{\approx} \pi/2 +  \sqrt{\varepsilon} \ .
\label{reg2}
\end{equation}
   
A seeming alternative to (\ref{Lreg1}) would be to introduce 
a regulator $\rho_{\varepsilon}(z)$ that would scale as a function 
of $z\!\equiv\!w/(a g)$ instead of $w$. But this possibility must be 
discarded because then the condition of suppressing the
pole structure [$\rho_{\varepsilon}(z) \ll 1$ for
$z \geq \pi/(ga)\!\equiv\!\pi/(m^2 {\tilde a})$]
cannot be reconciled with the condition of the
survival of a large chunk of the perturbative region
[$\rho_{\varepsilon}(z) \approx 1$ for
$z \leq \pi/(2 m^2 {\tilde a})$] 
at various values of ${\tilde a}$ simultaneously.

The conditions (\ref{reg2}) are designed in such a way
that the limit $\varepsilon\!\to\!+0$ would
apparently lead to the abrupt IR regulator appearing in
${\rm Re} \delta {\tilde {\cal L}}_0 ({\tilde a};p)$ 
of (\ref{L0}), with the abrupt cutoff at $w\!=\!\pi/2$.
We can choose the following specific one--parameter
family of regulators $\rho_{\varepsilon}(w)$
satisfying the afore--mentioned conditions:
\begin{eqnarray}
\rho_{\varepsilon}(w) &=& 
\frac{{\tilde \rho}_{\varepsilon}(w)}
{{\tilde \rho}_{\varepsilon}(0)} \ ,
\quad 
{\tilde \rho}_{\varepsilon}(w) = \frac{1}{2}
- \frac{1}{\pi} \arctan \left( 
\frac{ w - \pi/2}{\varepsilon} \right) \ .
\label{tildrho}
\end{eqnarray}
When $\varepsilon\!\to\!+0$ ($\varepsilon\!\not=\!0$),
these regulators differ from the abrupt cutoff regulator
outside the narrow $w$--interval $[\pi/2 - \sqrt{\varepsilon},
\pi/2 + \sqrt{\varepsilon}]$ by at most $\sim$$\sqrt{\varepsilon}$  
\begin{equation}
\rho_{\varepsilon}(w) =
\left \{ 
\begin{array}{l l}
1 - ({\varepsilon}/\pi)(\pi/2 - w)^{-1} + {\cal O}(\varepsilon^2) & 
\text{if $w < \frac{\pi}{2}\!-\!\sqrt{\varepsilon}$} \ , \\ 
({\varepsilon}/\pi)(w - \pi/2)^{-1} + {\cal O}(\varepsilon^2) &
\text{if $w > \frac{\pi}{2}\!+\!\sqrt{\varepsilon}$} \ ,
\end{array}
\right \} 
\label{rholim}
\end{equation}
while they may differ from the abrupt version significantly
only in the afore--mentioned narrow interval.
The first thing to check would be that the regularized 
expression (\ref{Lreg1}) with the regulator
(\ref{tildrho})--(\ref{rholim}), in the limit
$\varepsilon\!\to\!+0$ ($\varepsilon\!\not=\!0$)
really gives numerically the result (\ref{L0})
of the abrupt cutoff. Stated otherwise, we should
check that the $\lim_{\varepsilon \to +0}$ in front of
the integral (\ref{Lreg1}) can be moved into the
integral, without changing the result. 
For such a check, we need to
see that the contributions in (\ref{Lreg1}) from
the singular (poles) regions ($w > \pi/2$) are
suppressed toward zero when $\varepsilon\!\to\!+0$.
Such a check is straightforward and we performed it.
It turns out that the $w$--regions 
$[ (n\!-\!1/2) \pi, (n\!+\!1/2) \pi]$ 
around the $n$'th pole $w_n\!=n \pi$
are suppressed by a factor $\sim$$\varepsilon$
when $n \geq 2$, and by at least a factor
$\sim$$\sqrt{\varepsilon}$ when $n\!=\!1$.
Thus, all these contributions go to zero
when $\varepsilon\!\to\!+0$. On the other hand, 
on the $w$--interval $[0, \pi/2]$, there are no
singularities of the integrand and the regulator is virtually equal
to $1$ in the entire interval when $\varepsilon\!\to\!+0$.
Therefore, on this interval we can automatically push the limiting 
procedure into the integral.
Thus we really have
\begin{equation}
\lim_{\varepsilon \to +0} 
{\rm Re} \delta {\tilde {\cal L}}^{({\rm P.\/})}_{\varepsilon}
({\tilde a}; p) = 
{\rm Re} \delta {\tilde {\cal L}}_0({\tilde a};p) \ ,
\label{limeq}
\end{equation}
i.e., the numerical value of the perturbative part
with the infinitesimally ``softened'' IR cutoff is the same 
as that of the perturbative part with the abrupt
IR cutoff (\ref{L0}).

Now we will investigate the expansions of the above
two expressions around the point ${\tilde a}\!=\!0$,
in order to see the difference in the (non)analiticity
structure between the two cases. 
We can find the small--${\tilde a}$ expansion of
${\rm Re} \delta {\tilde {\cal L}}_0({\tilde a};p)$
of (\ref{L0}) by expanding first the integrand 
(without the exponent) there, i.e., the Borel transform, 
in powers of $w$. As argued in Section IV [cf. Eqs.
(\ref{BL})--(\ref{EH4})], this expansion yields
(\ref{BL}) with ${\tilde a} \mapsto w$, where $c_j(p)$'s
are given by (\ref{cjs}). Then the term--by--term integration
over $w$ leads to the small--${\tilde a}$
expansion of ${\rm Re} \delta {\tilde {\cal L}}_0$ 
\begin{eqnarray}
\lefteqn{
{\rm Re} \delta {\tilde {\cal L}}_0({\tilde a};p)^{({\rm exp.\/})}
 =  c_1(p) \int_0^{\pi/2} dw \exp(-w/{\tilde a}) w
+ c_3(p) \int_0^{\pi/2} dw \exp(-w/{\tilde a}) w^3 +
\cdots
}
\nonumber\\
&=& \left[ c_1(p) 1!\,{\tilde a}^2 + c_3(p) 3!\,{\tilde a}^4 
+ \cdots \right] - 
{\tilde a} \exp \left(- \frac{\pi}{2 {\tilde a}} \right) \left[
c_1(p) \left( \frac{\pi}{2} \right) +
c_3(p) \left( \frac{\pi}{2} \right)^3 + \cdots \right]
\nonumber\\
&&+ {\cal O} \left( {\tilde a}^2 \exp[ - \pi/(2 {\tilde a})] \right) \ .
\label{L0exp}  
\end{eqnarray}
Incidentally, the coefficient at 
${\tilde a} \exp[- \pi/(2 {\tilde a})]$,
written as an infinite sum, is just the value of the Borel
transform at $w=\pi/2$ [cf. remark following Eq. (\ref{EH4})]
\begin{equation}
\left[
c_1(p) \left( \frac{\pi}{2} \right) +
c_3(p) \left( \frac{\pi}{2} \right)^3 + \cdots \right]
= \left( \frac{2}{\pi} \right) \left[
\left( \frac{2}{\pi} \right)^2 - \frac{1}{3} (1\!-\!p^2) \right] \ .
\label{coef1L0}
\end{equation}
The coefficients of terms 
${\cal O} \left( {\tilde a}^2 \exp[- \pi/(2 {\tilde a})] \right)$
can be obtained in an analogous manner, by using derivatives of
the Borel transform with respect to $w$ at $w\!=\!\pi/2$.
Expressions (\ref{L0exp})--(\ref{coef1L0}) show explicitly
the following: In the small--${\tilde a}$ expansion of 
${\rm Re} \delta {\tilde {\cal L}}_0$ of (\ref{L0}), 
in addition to the usual perturbation expansion part
(\ref{Lpert}) that is analytic at ${\tilde a}\!=\!0$,
we obtain formally also terms 
$\sim$${\tilde a}^n \exp[- \pi/(2 {\tilde a})]$
which are nonanalytic at ${\tilde a}=0$. 
One might suspect that such terms could
possibly be of nonperturbative origin, and below we
will show that they are not. More specifically, we will show that
they are an artifact of the abruptness of the IR cutoff
and that they are de facto not there, in the sense that
they disappear when we consider instead of
${\rm Re} \delta {\tilde {\cal L}}_0$ its numerical
equivalent, i.e., the $\varepsilon\!\to\!+0$ limit
of the left--hand side of (\ref{limeq}). 
To show this, we have to expand the latter expression 
[at $\varepsilon\!\not=\!0$ -- i.e. (\ref{Lreg1})]
around ${\tilde a}\!=\!0$. For that, we first Taylor--expand
the regulator $\rho_{\varepsilon}(w)$ (\ref{tildrho}),
which is analytic everywhere,\footnote{In contrast to the
abrupt cutoff when $\rho_0(w)\!=\!1$ for $w\!<\!(\pi/2)$,
and $\rho_0(w)\!=\!0$ for $w\!>\!(\pi/2)$.}
in powers of $w$ for small $\varepsilon$
\begin{eqnarray}
{\tilde {\rho}}_{\varepsilon}(w) & = &
{\tilde {\rho}}_{\varepsilon}(0) - 
w\,\frac{\varepsilon}{2} \left( \frac{2}{\pi} \right)^3
- \cdots - w^n\;\frac{\varepsilon}{2} \left( \frac{2}{\pi} \right)^{n+2}
- \cdots + {\cal O}(\varepsilon^3) \ ,
\label{trhoexp}
\\
{\tilde {\rho}}_{\varepsilon}(0) & = & 1 - \frac{2}{\pi} {\varepsilon}
+ {\cal O}(\varepsilon^3) \ .
\label{trho0}
\end{eqnarray}
The other part of the integrand in (\ref{Lreg1}), without the
exponent, is the Borel transform whose small--${\tilde a}$ 
expansion is (\ref{BL}) with ${\tilde a} \mapsto w$. 
Combining this and (\ref{trhoexp})--(\ref{trho0}), we obtain
after some straightforward algebra\footnote{
We again integrate term--by--term; and we repeatedly use the identity: 
$\int_0^{\infty} du \exp(-u) u^n = n!$.}
the small--${\tilde a}$ expansion of
(\ref{Lreg1}) around ${\tilde a}\!=\!0$ for small $\varepsilon$
\begin{eqnarray}
\lefteqn{
{\rm Re} \delta {\tilde {\cal L}}^{({\rm P.\/})}_{\varepsilon}
({\tilde a};p)^{({\rm exp.\/})} = 
\left[ c_1(p) 1!\,{\tilde a}^2 + c_3(p) 3!\,{\tilde a}^4 
+ c_5(p) 5!\,{\tilde a}^6 \cdots \right]
}  
\nonumber\\
&& - \varepsilon \frac{1}{2} \left( \frac{2}{\pi} \right)^3
{\Bigg \{} c_1(p) 2!\,{\tilde a}^3 
+ \left( \frac{2}{\pi} \right) c_1(p) 3!\,{\tilde a}^4 
\nonumber\\
&&
+ \left[ \left( \frac{2}{\pi} \right)^2 c_1(p) +
c_3(p) \right] 4!\,{\tilde a}^5 +
\left[ \left( \frac{2}{\pi} \right)^3 c_1(p) +
\left( \frac{2}{\pi} \right) c_3(p) \right] 5!\,{\tilde a}^6
\nonumber\\
&&
+ \left[ \left( \frac{2}{\pi} \right)^4 c_1(p) +
\left( \frac{2}{\pi} \right)^2 c_3(p) + c_5(p) \right] 6!\,{\tilde a}^7
+ \cdots {\Bigg \}} + {\cal O}(\varepsilon^2) \ .
\label{Lreg1exp}
\end{eqnarray}
Here we see explicitly that the small--${\tilde a}$ expansion of
${\rm Re} \delta {\tilde {\cal L}}^{({\rm P.\/})}_{\varepsilon}
({\tilde a};p)$ of (\ref{Lreg1}) at nonzero $\varepsilon$
exists and that this function is analytic there,
having no nonanalytic terms 
$\sim$$\exp(- {\rm const.\/}/{\tilde a})$,
in contrast to the expansion of
${\rm Re} \delta {\tilde {\cal L}}_0$ 
where $\varepsilon$ was set equal to zero exactly 
(i.e., inside the integral). 
Further, expansion (\ref{Lreg1exp}) goes over into the
usual perturbation expansion 
$\delta {\tilde {\cal L}}^{\rm pert.}$ (\ref{Lpert})
when $\varepsilon\!\to\!+0$.

These considerations thus lead us to the following
conclusions:
\begin{itemize}
\item
The perturbative part of the
induced Lagrangian density,
${\rm Re} \delta {\tilde {\cal L}}_0$
as defined in (\ref{L0}),
has an abrupt IR cutoff at $w\!=\!\pi/2$,
and it is numerically equal to the
corresponding expression with an infinitesimally
softened IR cutoff -- cf. left--hand side
of (\ref{limeq}).
\item
The small--${\tilde a}$ expansion of
${\rm Re} \delta {\tilde {\cal L}}_0({\tilde a}; p)$
reproduces the usual perturbation expansion (\ref{Lpert})
plus nonanalytic terms
$\sim$${\tilde a}^n \exp(- {\rm const.\/}/{\tilde a})$
[cf. (\ref{L0exp})].
\item
The small--${\tilde a}$ expansion of the corresponding expression
(\ref{Lreg1}) with a softened IR cutoff
($\varepsilon\!\not=\!0$) yields no nonanalytic terms; when the
softening of the IR cutoff becomes infinitesimal
($\varepsilon\!\to\!+0$, $\varepsilon\!\not=\!0$),
the expansion becomes identical with that
of the usual perturbation expansion (\ref{Lpert}).
\item
The above points show that the nonanalytic terms
in the small--${\tilde a}$ expansion of 
${\rm Re} \delta {\tilde {\cal L}}_0({\tilde a}; p)$
are only an artifact of the abruptness of the
IR cutoff (the cutoff regulator becomes a
nonanalytic function of the proper time $w$)
and are thus not of a nonperturbative physical
origin. ${\rm Re} \delta {\tilde {\cal L}}_0({\tilde a}; p)$
should be reinterpreted as the limit with
the infinitesimally softened IR cutoff
[the left--hand side of (\ref{limeq})],
the latter being numerically the same
but its small--${\tilde a}$ expansion having
no nonanalytic terms.
\end{itemize}

\end{appendix}

\newpage

\noindent

\noindent

\noindent

\end{document}